\documentclass[fleqn,usenatbib,useAMS]{mnras}

\usepackage{graphicx}	
\usepackage{amsmath}	
\usepackage{amssymb}	
\usepackage{multicol}        
\usepackage{bm}		
\usepackage{pdflscape}	

\usepackage[T1]{fontenc}
\usepackage{ae,aecompl}

\usepackage{newtxtext,newtxmath}

\newcommand{\hbeta}{H{$\beta$}}
\newcommand{\halpha}{H{$\alpha$}}

\def\CIV{C\,{\sc iv}}
\def\MgII{Mg\,{\sc ii}}
\def\FeII{Fe\,{\sc ii}}
\def\NaI{Na\,{\sc i}}
\def\CaII{Ca\,{\sc ii}}

\title[Small-scale double quasars in DECaLS]{Small-scale double quasars discovered in the DECam Legacy Survey}

\author[Chen]{Yu-Ching Chen$^{1,2,3}$\thanks{Contact e-mail: \href{mailto:ycchen@illinois.edu}{ycchen@illinois.edu}}
\\
$^{1}$Department of Astronomy, University of Illinois at Urbana-Champaign, 1002 West Green Street, Urbana, IL 61801, USA\\
$^{2}$National Center for Supercomputing Applications, 1205 West Clark Street, Urbana, IL 61801, USA\\
$^{3}$Center for AstroPhysical Surveys, National Center for Supercomputing Applications, Urbana, IL, 61801, USA\\
}

\date{Accepted XXX. Received YYY; in original form ZZZ}

\pubyear{2021}

\begin{document}
\label{firstpage}
\pagerange{\pageref{firstpage}--\pageref{lastpage}}
\maketitle

\begin{abstract}
Dual quasars are precursors of binary supermassive black holes, which are important objects for gravitational wave study, galaxy evolution, and cosmology. I report four double quasars, one of which is newly discovered, with separations of 1--2 arcsec selected from the DECam Legacy Survey. The Gemini optical spectra confirm that both sources are quasars at the same redshifts. J0118-0104 and J0932+0722 are classified as dual quasars, whereas J0037+2058 could be either a dual quasar or a lensed quasar. I estimate the physical properties such as black hole masses and Eddington ratios for the four double quasars. The newly discovered system supplements the incomplete sample of dual quasars at $\sim$10 kpc at $z>0.5$. The results demonstrate that new high-quality imaging surveys with existing spectroscopic data could reveal additional small-scale double quasars. Combined with the dual quasars from the literature, I find a power-law index of 1.45$\pm$0.48 for the distribution of dual quasars as a function of separation between 5--20 kpc, steeper than that expected from dynamical friction, though it is likely due to the incomplete sample at $<$15 kpc.
I also discuss the implications for future searches with new imaging surveys such as the Dark Energy Survey and the Legacy Survey of Space and Time. 
\end{abstract}





\newpage

\section{Introduction}

Observational evidences support that most massive galaxies contain supermassive black holes \citep[SMBHs; ][]{kormendy95,KormendyHo2013}. Thus, galaxy mergers should lead to the formation of binary supermassive black holes \citep[BSBHs; ][]{begelman80}. BSBHs are important objects for gravitational wave study, galaxy evolution and cosmology. While the formation of BSBHs is inevitable, direct evidence has so far been elusive. There are only two candidates BSBHs with separation of pc scales that were directly imaged by the Very Long Baseline Array \citep{Rodriguez2006,Kharb2017} and hundreds of candidates discovered by the indirect methods \citep[such as periodic light curves and radial velocity shifts,][]{eracleous11,Shen2013,Liu2014,Graham2015,Charisi2016,Runnoe2017,Wang2017,Guo2019,LiuT2019,ChenYC2020,Liao2021}.

In the past decade, thanks to large spectroscopic surveys such as the Sloan Digital Sky Survey, significant progress has been made in finding dual quasars, the progenitors of the BSBHs when they are active and still evolving in the potential of the (merged) host galaxy \citep{Liu11,Hennawi06,Hennawi10,Eftekharzadeh17}. However, the angular diameter distance increases dramatically from local universe ($z=0$) to high-redshift universe ($z=2$), It is difficult to identify close dual quasars systematically in optical surveys due to the angular resolution limits of ground-based telescopes. Only handful dual quasars with projected separation $r_p\lesssim$10 kpc at redshift $z>0.5$ are discovered  \citep{Junkkarinen01,Inada08,Inada12,More16,Schechter17,Anguita18,Lemon18,Lemon20,Silverman20,Shen2021}.

New imaging surveys with excellent image quality and improved classification pipeline enable the discovery of small-separation double quasars that are not found before. In the paper, I observe eight close (1--2 arcsec) double quasar candidates selected from DECaLS imaging survey with Gemini and report four spectroscopically confirmed double quasars. Among the four confirmed double quasars, one is newly discovered and the other three have been reported \citep{Inada08,More16,Hutsemekers2020,Lemon20}. The paper is organized as follows. In \autoref{sec:data}, I describe the data and the target selection for this study. I present the main results including DECaLS images and Gemini/GMOS optical spectra and classify the individual systems in \autoref{sec:results}. The implications are discussed in \autoref{sec:discussions}. I summarize the findings in \autoref{sec:conclusions}. All physical separations are the projected separation. A flat $\Lambda$CDM cosmology is adopted throughout with $\Omega_\Lambda=0.7$, $\Omega_m$=0.3, and $H_0=70\,{\rm km\,s^{-1}Mpc^{-1}}$.

\section{Data and Target selection}
\label{sec:data}
\subsection{The Dark Energy Camera Legacy Survey (DECaLS)}
The Dark Energy Camera Legacy Survey (DECaLS) is parts of the DESI Legacy Imaging Surveys, which will provide optical imaging in $grz$ filters for targets in two third of DESI footprint \citep{DECaLS}. DECaLS uses the Dark Energy Camera \citep[DECam,][]{DECam} on the Blanco 4m telescope, located at the Cerro Tololo Inter-American Observatory. DECaLS can reach the median 5$\sigma$ detection limit of 23.95, 23.54, 22.50 in the $grz$ filters, respectively. The delivered image quality is 1.3, 1.2, and 1.1 arcsec in $grz$, respectively \citep{DECaLS}. The double quasars candidates are selected from the DECaLS data release 8, which was available in February 2019. 

\subsection{Double quasar candidates}
\begin{figure}
	\includegraphics[width=0.98\columnwidth]{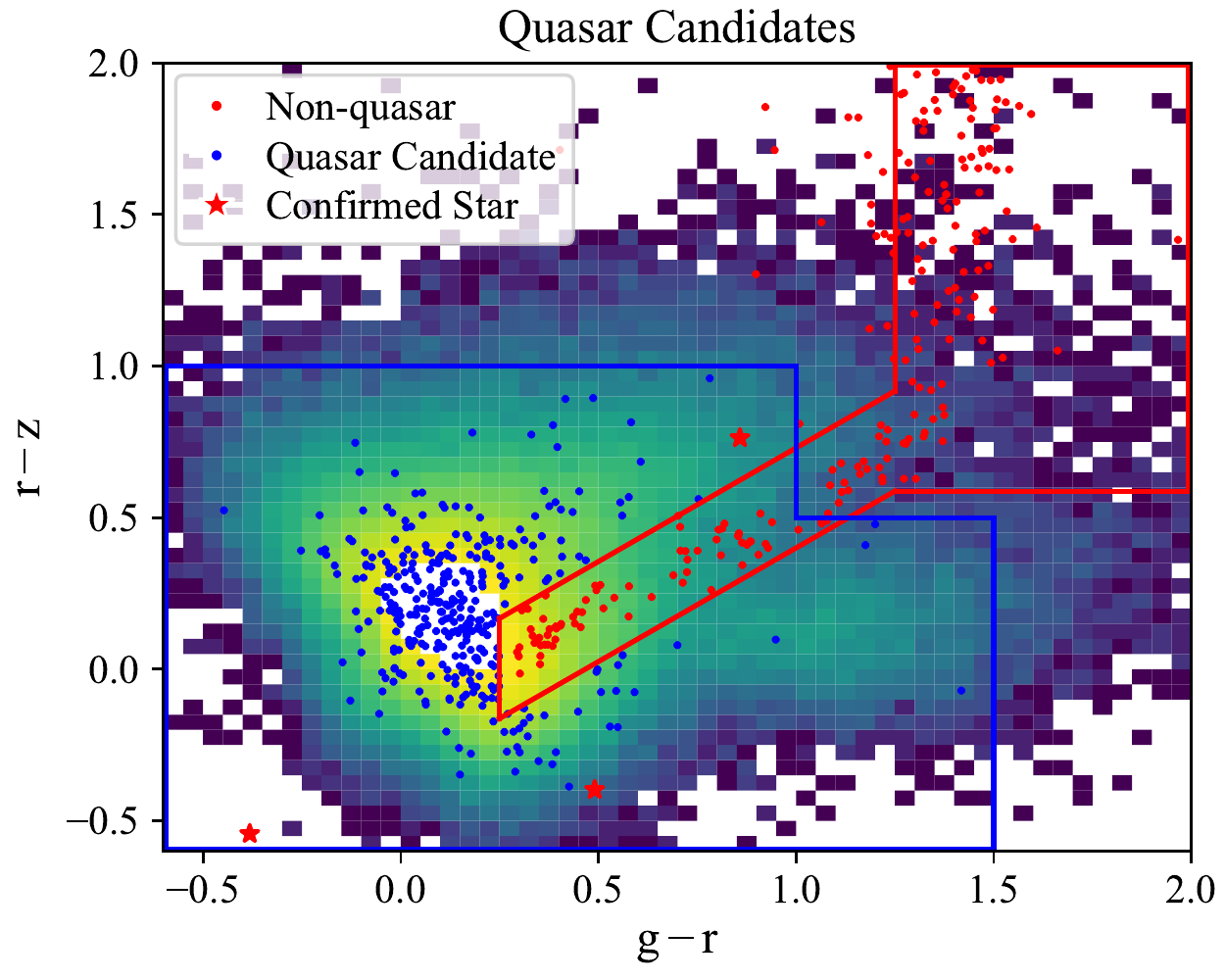}
	\includegraphics[width=0.98\columnwidth]{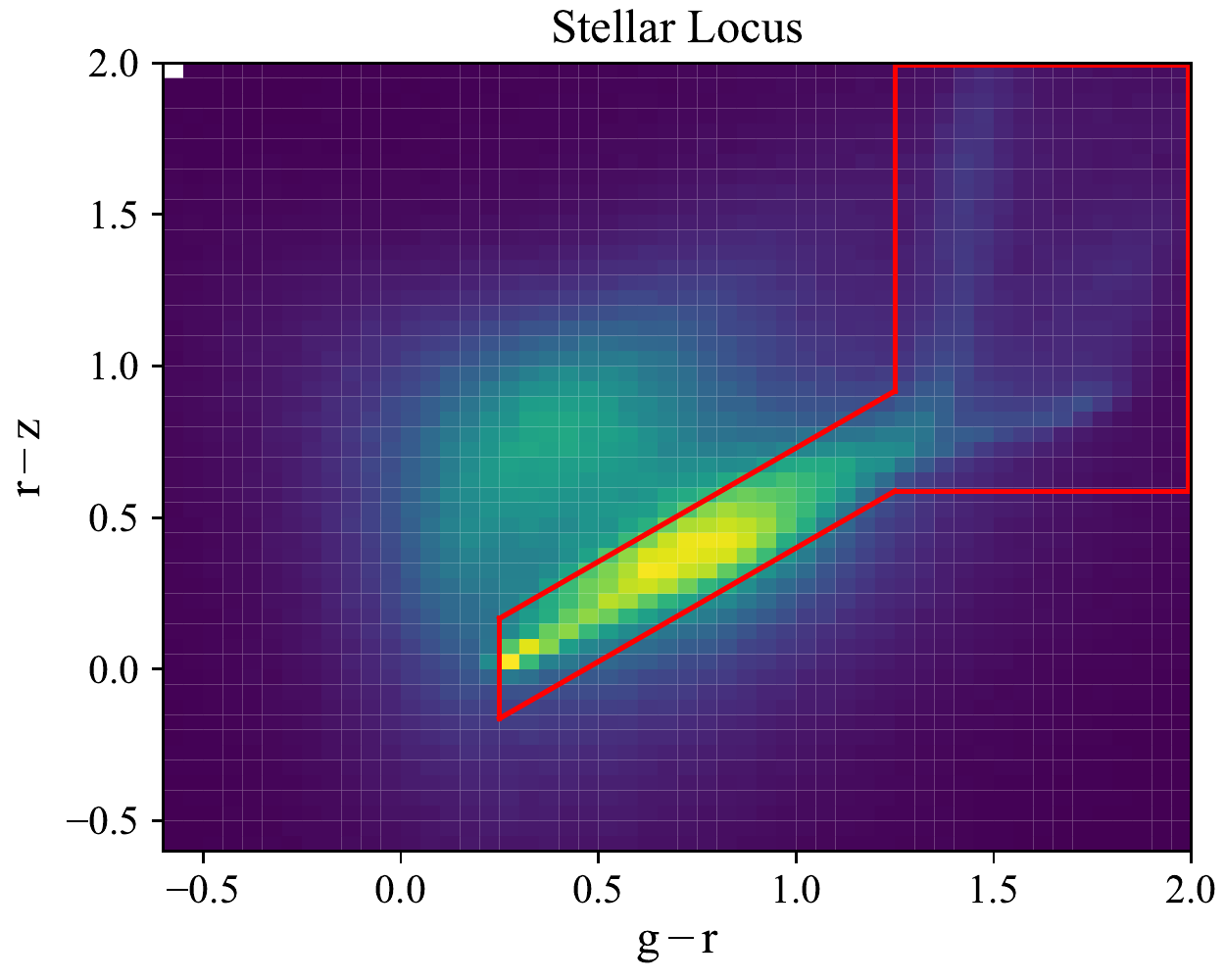}
    \caption{Color-color ($g-r$ vs. $r-z$) diagrams of the double quasar candidates and the color selection criteria. {\it Top}: Colors of each source of the 304 quasars having companions with separation of 1--2 arcsec in DECaLS. The 2D color map is the color distribution of the SDSS DR14 spectroscopically confirmed quasars \citep{Paris2018}. The source is considered as double quasar candidate if both cores are inside the blue boundary, where most quasars are located, and are outside the red boundary, which represents the stellar locus. The red stars are the three stars confirmed in the follow-up spectra. {\it Bottom}: Color distribution of all DECaLS sources at region where Right Ascension between 36 deg and 43 deg and Declination between -1.25 deg and 1.25 deg. The red boundary marks out the stellar locus.}
    \label{fig:color}
\end{figure}

The parent sample is selected from the Sloan Digital Sky Survey (SDSS) Data Release (DR) 14 quasar catalog \citep{Paris2018}. I search close companions in DECaLS around the 281,006 spectroscopically confirmed quasars; 307 quasars have companions with separation of 1--2 arcsec inside the DECaLS footprint. To avoid the chance superposition with foreground stars, I then apply the color selection to the candidates, requiring both sources have color similar to the SDSS quasars and not in the stellar locus. \autoref{fig:color} shows the distribution of the candidates in the color-color ($r-z$ vs. $g-r$) diagram and the selection criteria. The target is considered as a double quasar candidate if the colors of both sources are consistent with typical SDSS quasars and are not inside the stellar locus (see \autoref{fig:color} for details). Though the selection might remove quasars with similar colors as stars, I prefer a small cleaner sample than a larger mixed bag for this pilot study. There are 34 targets passing the color selection. To avoid duplicate observations with the spectroscopically confirmed cases, I further removed the candidates which are confirmed lensed quasars, quasar/star pairs, and quasar pairs with different redshifts \citep{Inada08,Inada10,Kayo10,Inada12,Jackson12,More16}. 3 candidates with $r$-band magnitude $>22$ are removed due to faintness. The final sample consists of 13 double quasars candidates.

\subsection{Gemini spectroscopy}
I obtain optical long-slit spectroscopy with Gemini Multi-Object Spectrographs (GMOS) for 8 out of the 13 candidates (Program ID: GN-2020A-Q-232 and GN-2020B-FT-211; PI: Chen). Observation dates and total exposure time are listed in \autoref{tab:targets}. The R150--G5326 grating is used with a long-slit width of 0.75 arcsec. The GMOS spectra cover wavelength of 3600--10300 \AA~ with an effective spectral resolution $R\sim421$. The slits are oriented in the direction through both sources. The images are dithered in spatial direction and wavelength direction to avoid bad pixels and the CCD gaps. Two flux standard stars Wolf1346 and G191B2B are observed using the same configuration.

\begin{table*}
\addtolength{\tabcolsep}{-2pt}
 \caption{Properties of double quasar candidates and detail of the Gemini/GMOS observations. Listed from left to right are source names in J2000 coordinates, DECaLS AB magnitudes in $grz$ bands, separations of two sources, position angles between two sources in degree east of north, total exposure time of the GMOS observations, identifications of each source, classifications of each system, and references if objects are previously reported (1: \citealt{Lemon20}; 2:\citealt{More16}; 3:\citealt{Hutsemekers2020}; 4: \citealt{Inada08}). }
 \label{tab:targets}
 \begin{tabular}{ccccccccccccc}
  \hline\hline
  Object & R.A. & Decl. &  $m_g$ & $m_r$ & $m_z$ & Sep. & P.A. & Obs. Date & Exp. Time & Source & Classification & Ref.\\
  & (deg) & (deg) & & & & (arcsec) & (deg) & (UT) & (s) & & \\
  \hline
  J0037+2058a & 9.31620 & 20.97377 & 19.45 & 19.25 & 18.99 & 1.02 & 273.6 & 2021-01-05 & 960 & quasar ($z$=2.047) & dual/lensed \\
  J0037+2058b & 9.31590 & 20.97379 & 20.59 & 20.18 & 19.62&  &  &  &  & quasar ($z$=2.047) \\
  \hline
  J0118$-$0104a & 19.55014 & -1.07850 & 20.11 & 19.77 & 19.31 & 1.74 & 54.2 & 2021-01-07 & 960 & quasar ($z$=0.739) & dual & 1\\
  J0118$-$0104b & 19.55053 & -1.07821 & 20.44 & 20.41 & 20.38 &  &  &  &  & quasar ($z$=0.739) \\
  \hline
  J0818+0601a & 124.62696 & 6.02725 & 18.14 & 17.89 & 17.72 & 1.15 & 234.9 & 2020-02-03 & 1142 & quasar ($z$=2.363) & lensed & 2,3\\
  J0818+0601b & 124.62670 & 6.02706 & 20.03 & 19.80 & 19.45 & & & & & quasar ($z$=2.363)\\
  \hline
  J0932+0722a & 143.02982 & 7.38091 & 19.34 & 19.04 & 18.58 & 1.33 & 117.4 & 2020-02-04 & 3600 & quasar ($z$=1.996) & dual & 4\\
  J0932+0722b & 143.02949 & 7.38108 & 21.07 & 20.76 & 20.50 & & & & & quasar ($z$=1.989)\\
  \hline
  J1011$-$0302a & 152.92724 & -3.03704 & 18.90 & 18.73 & 18.57 & 1.99 & 100.0 & 2021-01-05 & 960 & quasar ($z$=0.859) & star-quasar \\
  J1011$-$0302b & 152.92779 & -3.03713 & 19.65 & 20.00 & 20.51 & & & & & F type star\\
  \hline
  J2147+2523a & 326.75677 & 25.38757 & 21.86 & 21.26 & 21.57 & 1.74 & 162.7 & 2020-07-02 & 4800 & quasar ($z$=1.175)  & star-quasar\\
  J2147+2523b & 326.75693 & 25.38711 & 21.92 & 21.67 & 21.50 & & & & & G type star\\
  \hline
  J2207+2522a & 331.87174 & 25.36786 & 19.97 & 19.45 & 19.76 & 1.29 & 40.3 & 2020-07-30 & 3600 & quasar ($z$=1.308) & star-quasar\\
  J2207+2522b & 331.87148 & 25.36759 & 21.44 & 20.49 & 19.64 & & & & & K type star\\
  \hline
  J2251+0016a & 342.94930 & 0.27795 & 19.97 & 19.63 & 19.11 & 1.87 & 148.6 & 2020-07-31 & 3600 & quasar ($z$=0.410) & projected quasars\\
  J2251+0016b & 342.94957 & 0.27751 & 21.24 & 20.93 & 20.42 & & & & & quasar ($z$=0.577)\\
  \hline
  \multicolumn{12}{l}{}
 \end{tabular}
\end{table*}

The standard Gemini {\sc iraf} routine is used to carry out bias subtraction, flat field correction, and cosmic ray rejection. The images are wavelength calibrated using the CuAr spectra taken at the same night and sky subtracted. Multiple images with same targets are combined to subtract the residual cosmic rays, bad pixels and CCD gaps. The combined 2-dimensional (2D) spectra are then flux calibrated using the flux standard stars. 

To spatially resolve close double sources, I collapse the 2D spectra along wavelength direction to construct the spatial profiles. The spatial profiles of the 8 targets are well described by the two Gaussian components with separations same as those measured from DECaLS imaging.
To account for image distortion and seeing change along wavelength direction, I first obtain the best-fit parameters including amplitudes of each source, full width at half maxima (FWHM) and the centroids of two Gaussian profiles with fixed distance at each wavelength. I then model the change of centroids and FWHM along the wavelength direction as a function of wavelength using a cubic spline. I finally decompose two sources at each wavelength to extract 1D spectra using the Gaussian components with a constant separation, and the wavelength-dependent centroids and FWHM derived from the spline fit.

\section{Results}
\label{sec:results}

\begin{figure*}
	\includegraphics[width=0.2\textwidth]{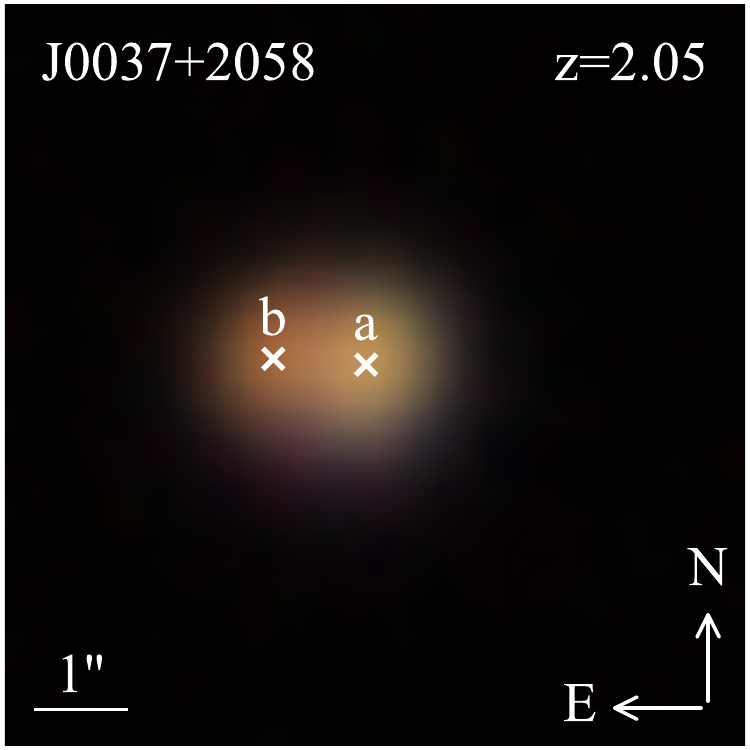}
	\includegraphics[width=0.2\textwidth]{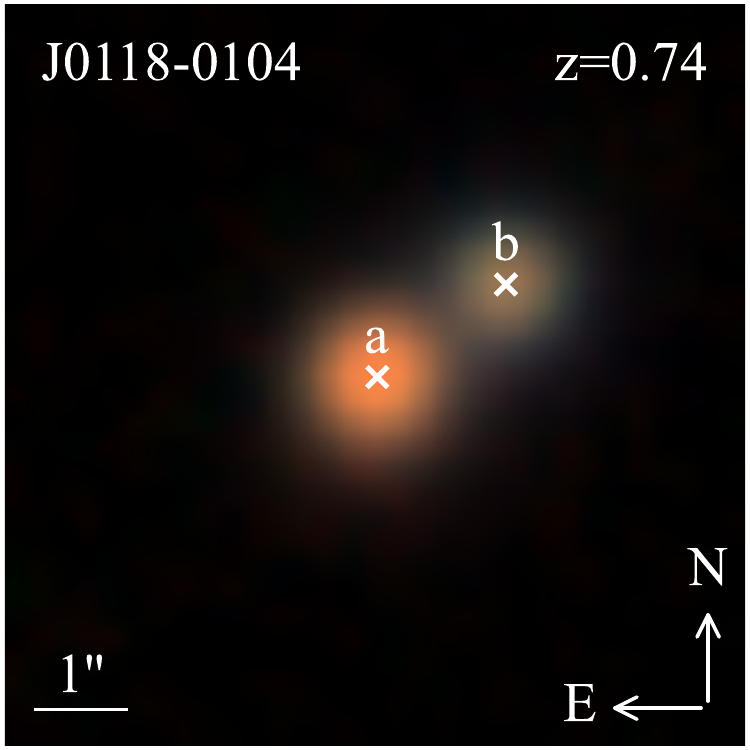}
	\includegraphics[width=0.2\textwidth]{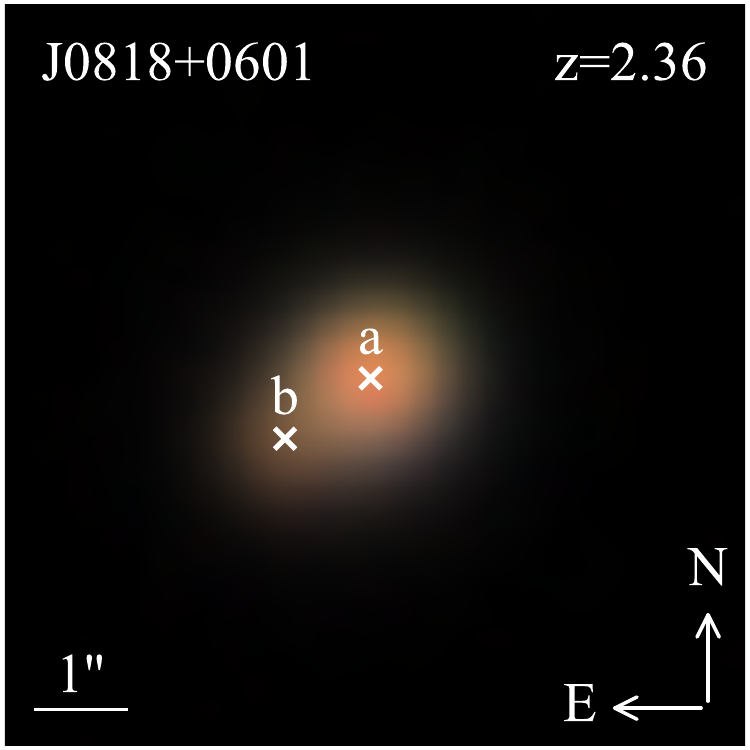}
	\includegraphics[width=0.2\textwidth]{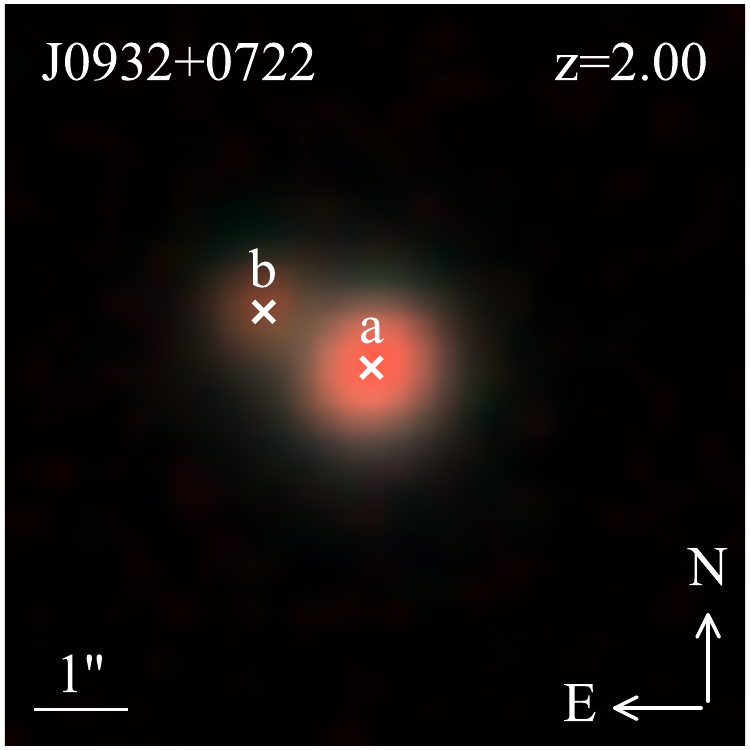}
	\includegraphics[width=0.2\textwidth]{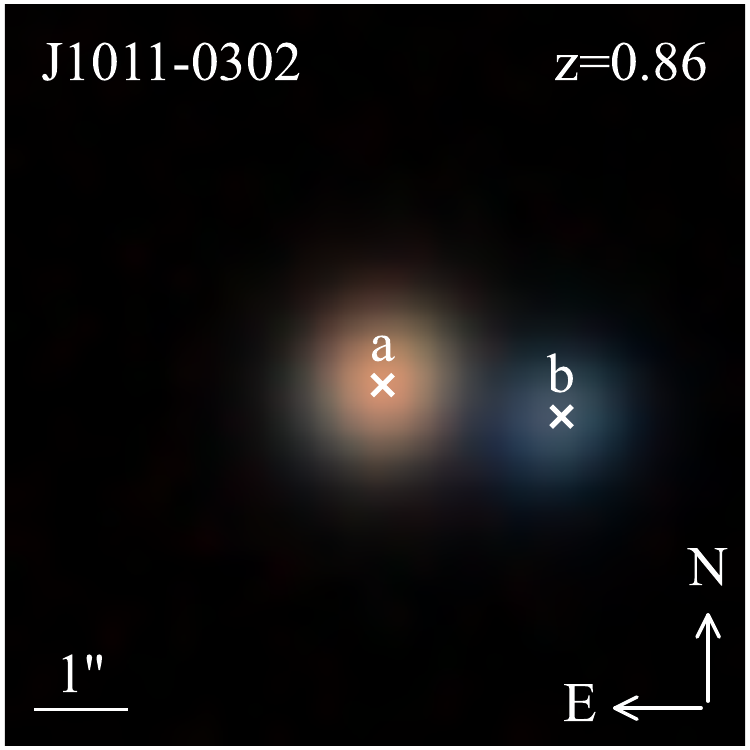}
	\includegraphics[width=0.2\textwidth]{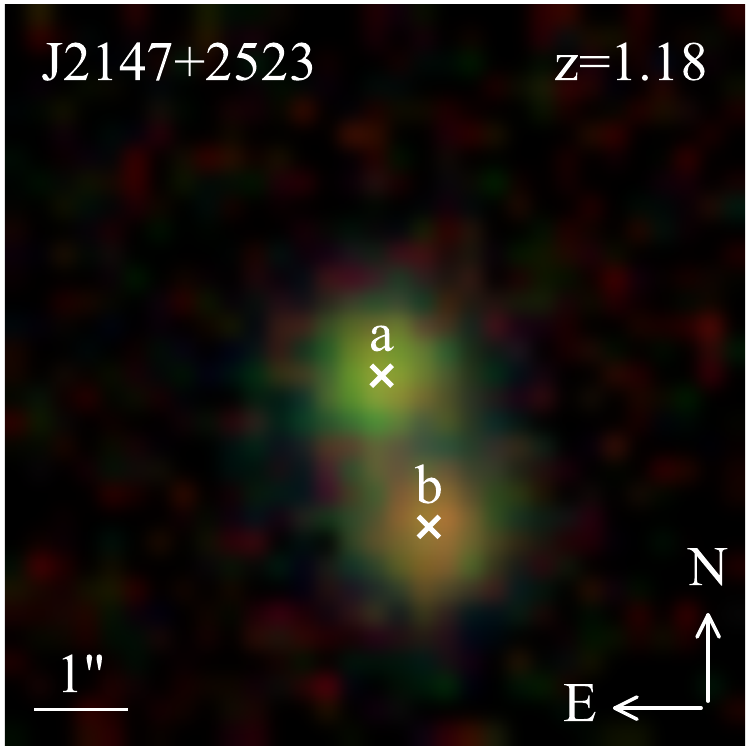}
	\includegraphics[width=0.2\textwidth]{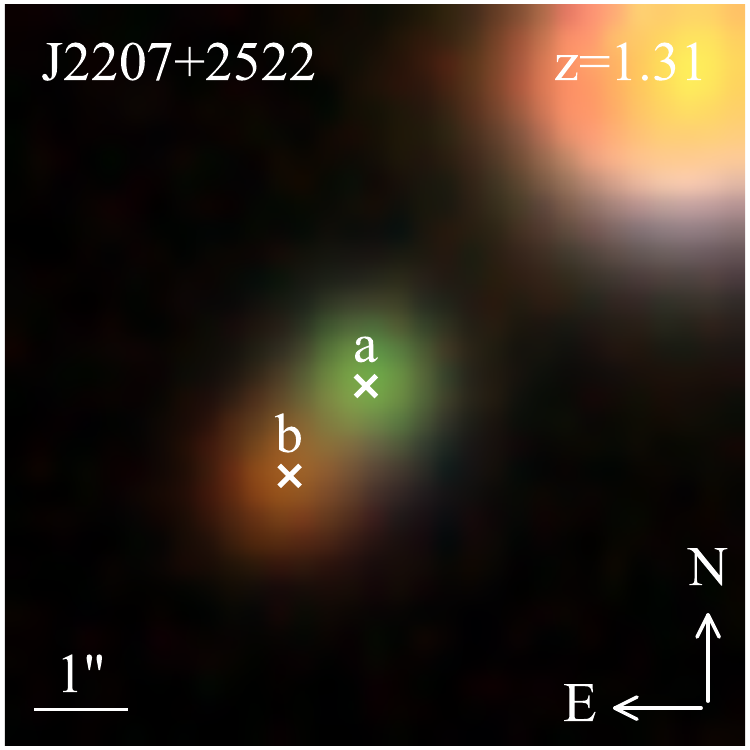}
	\includegraphics[width=0.2\textwidth]{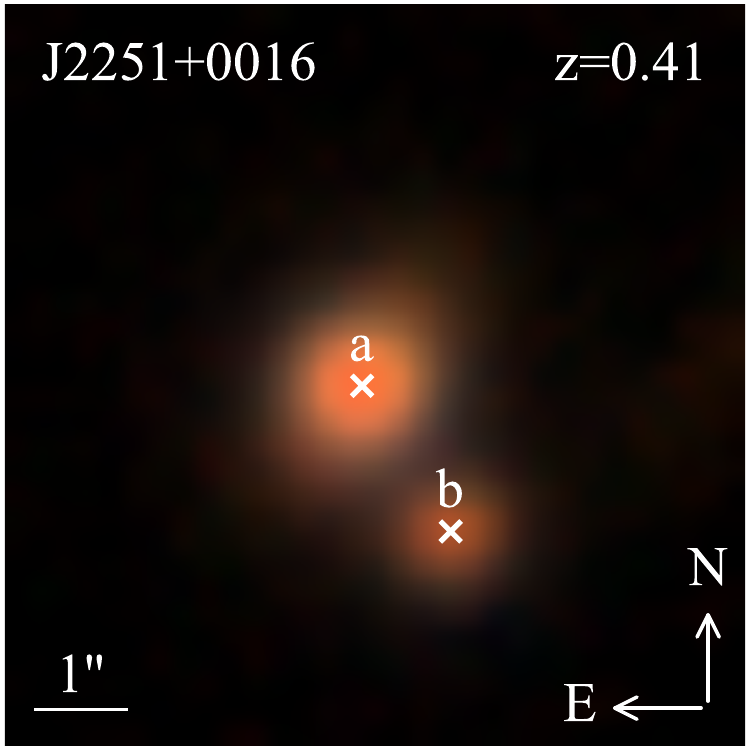}
	
    \caption{DECaLS $grz$ color composite images of the 8 double quasar candidates. The image size is 8 arcsec by 8 arcsec. North is up and east is to the left. The spectroscopic redshift is denoted at the top-right corner. The positions of each source are denoted by the white crosses. }
    \label{fig:decals_images}
\end{figure*}

\begin{figure*}
	\includegraphics[width=0.33\textwidth]{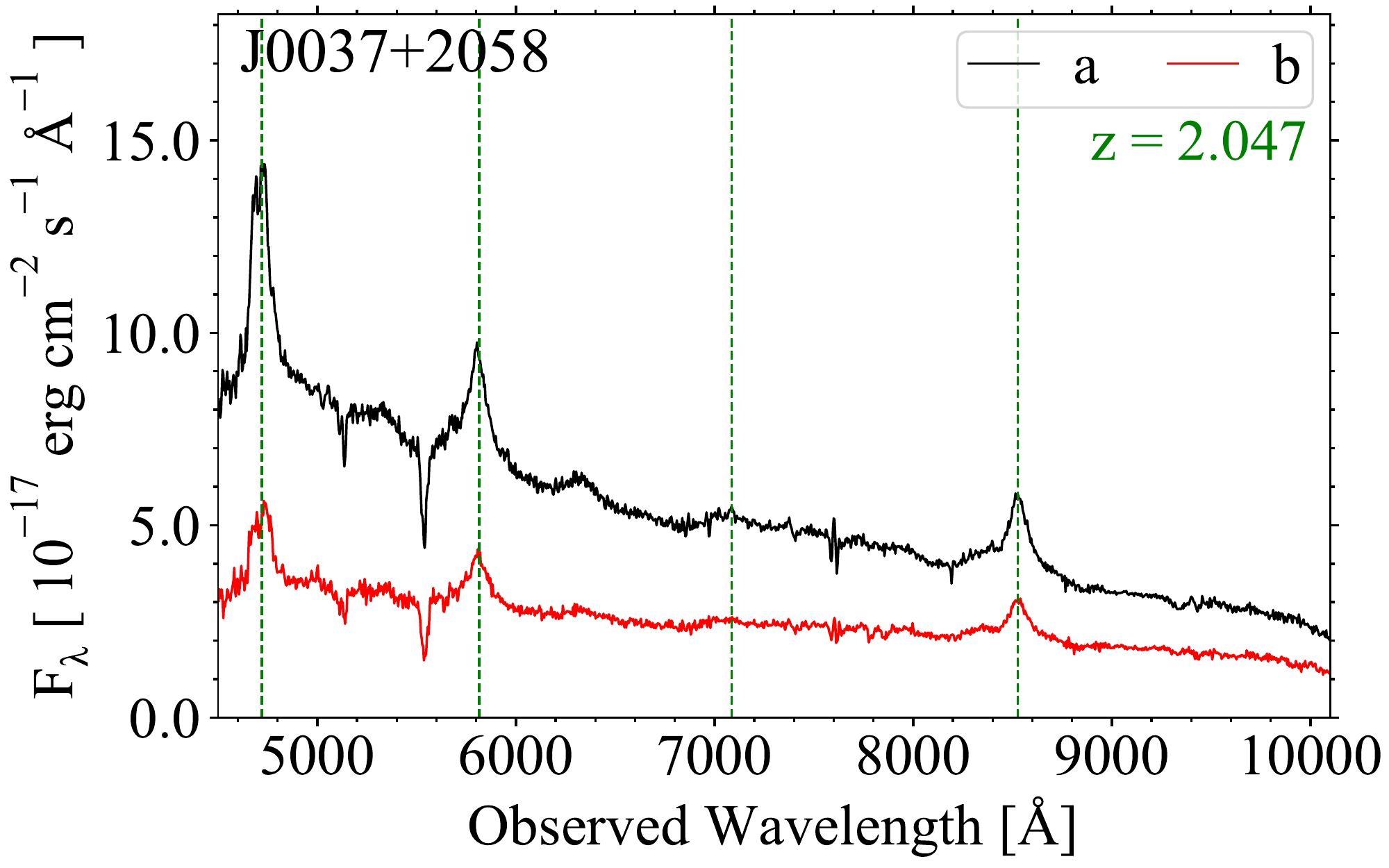}
	\includegraphics[width=0.33\textwidth]{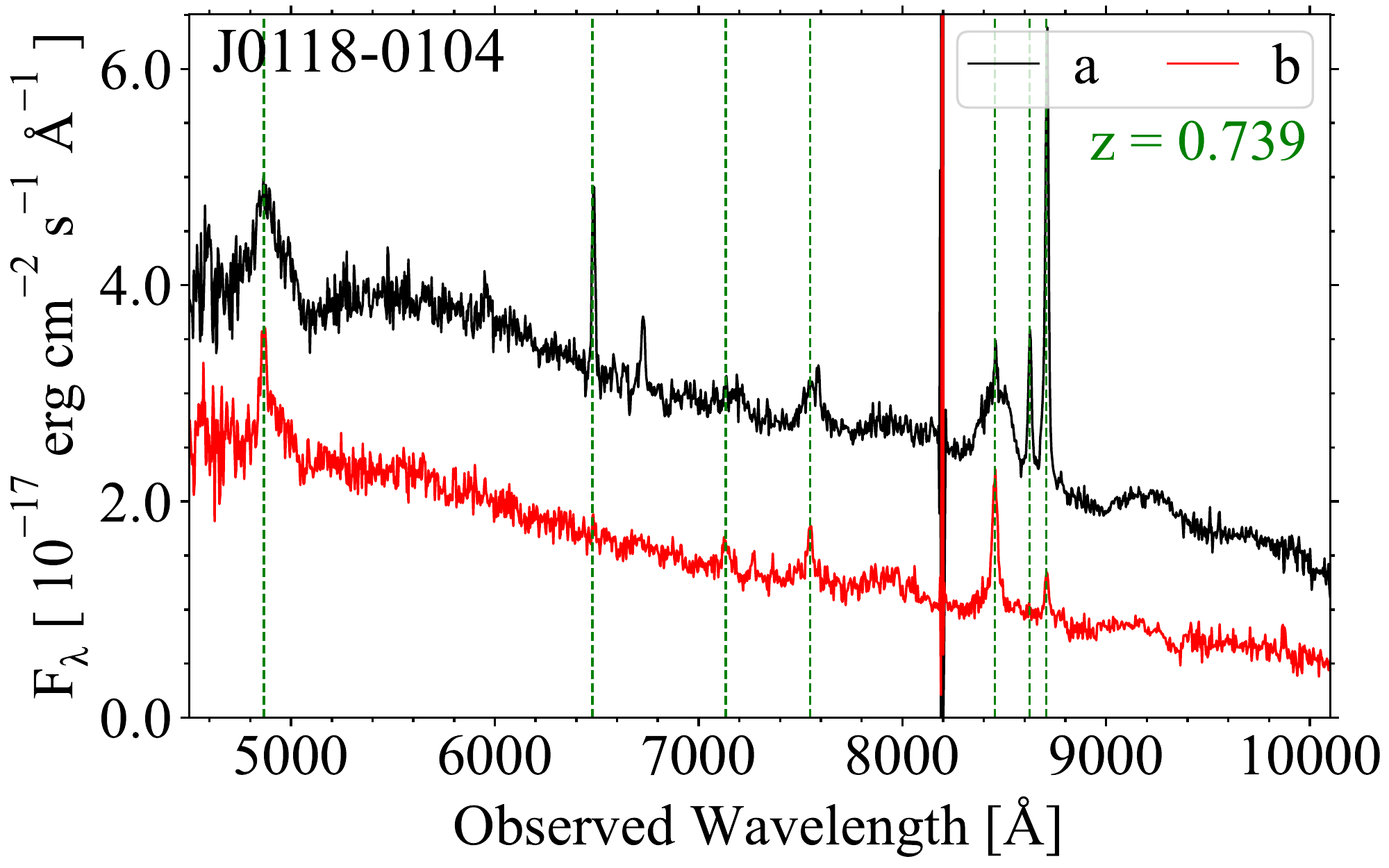}
	\includegraphics[width=0.33\textwidth]{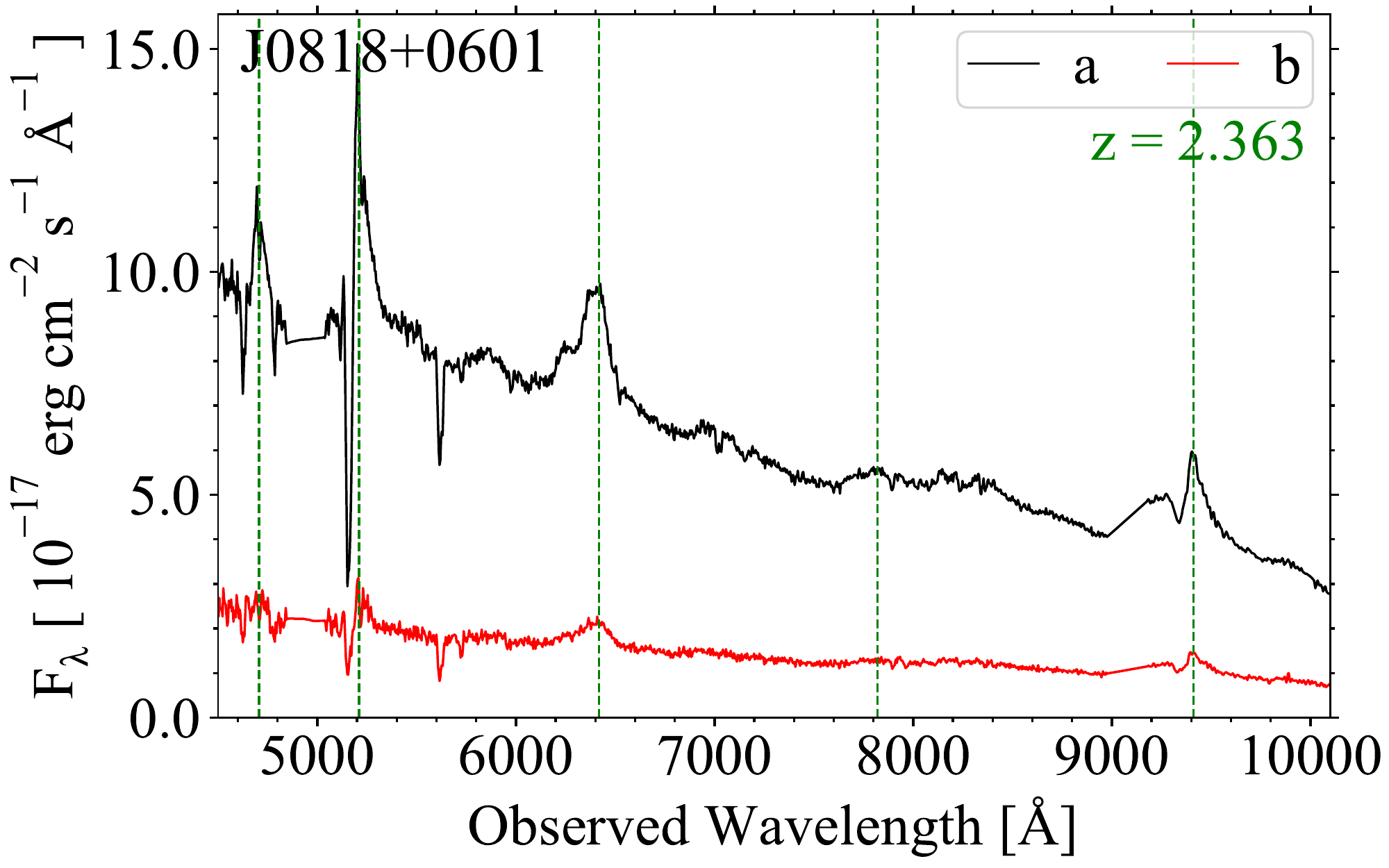}
	\includegraphics[width=0.33\textwidth]{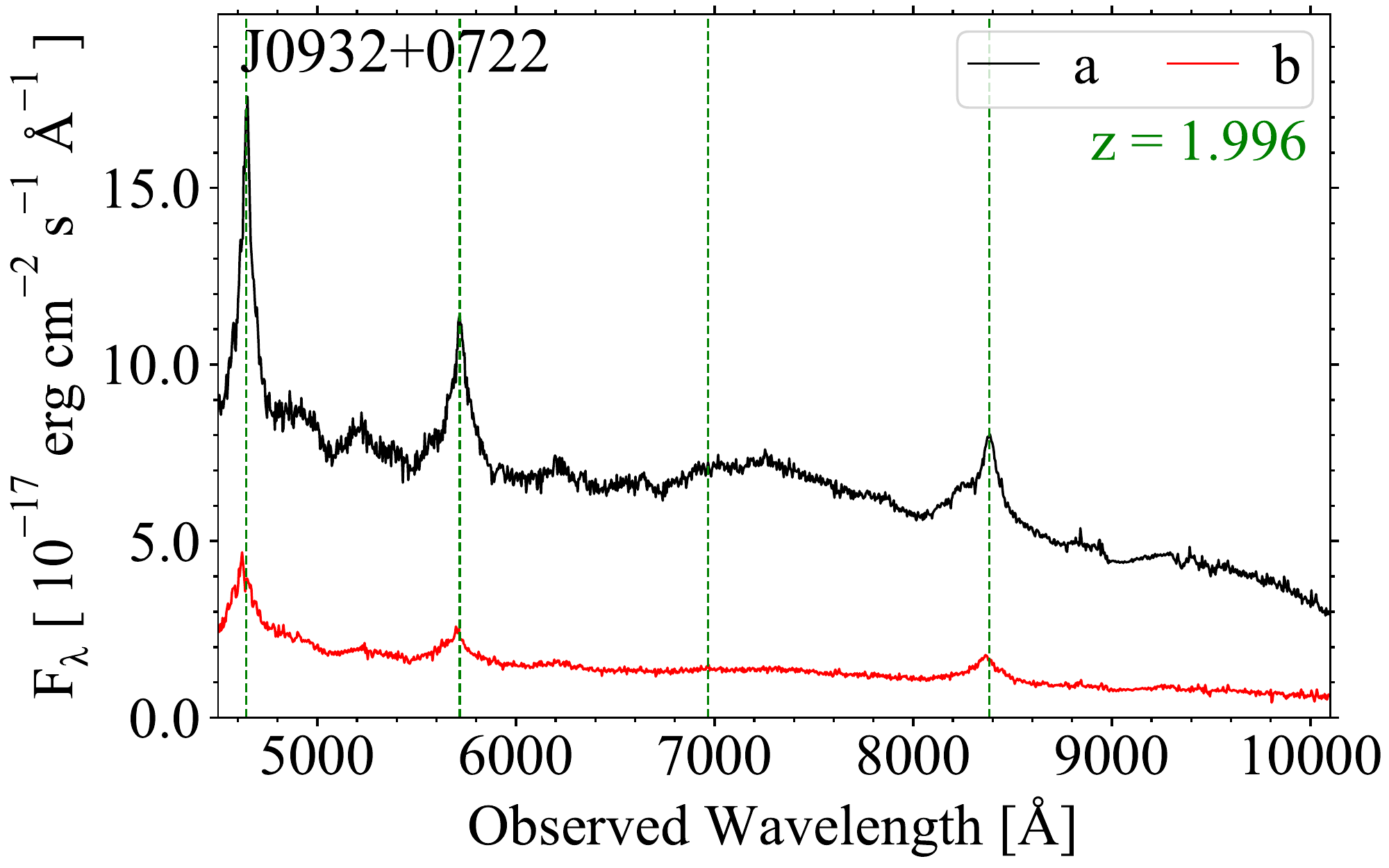}
	\includegraphics[width=0.33\textwidth]{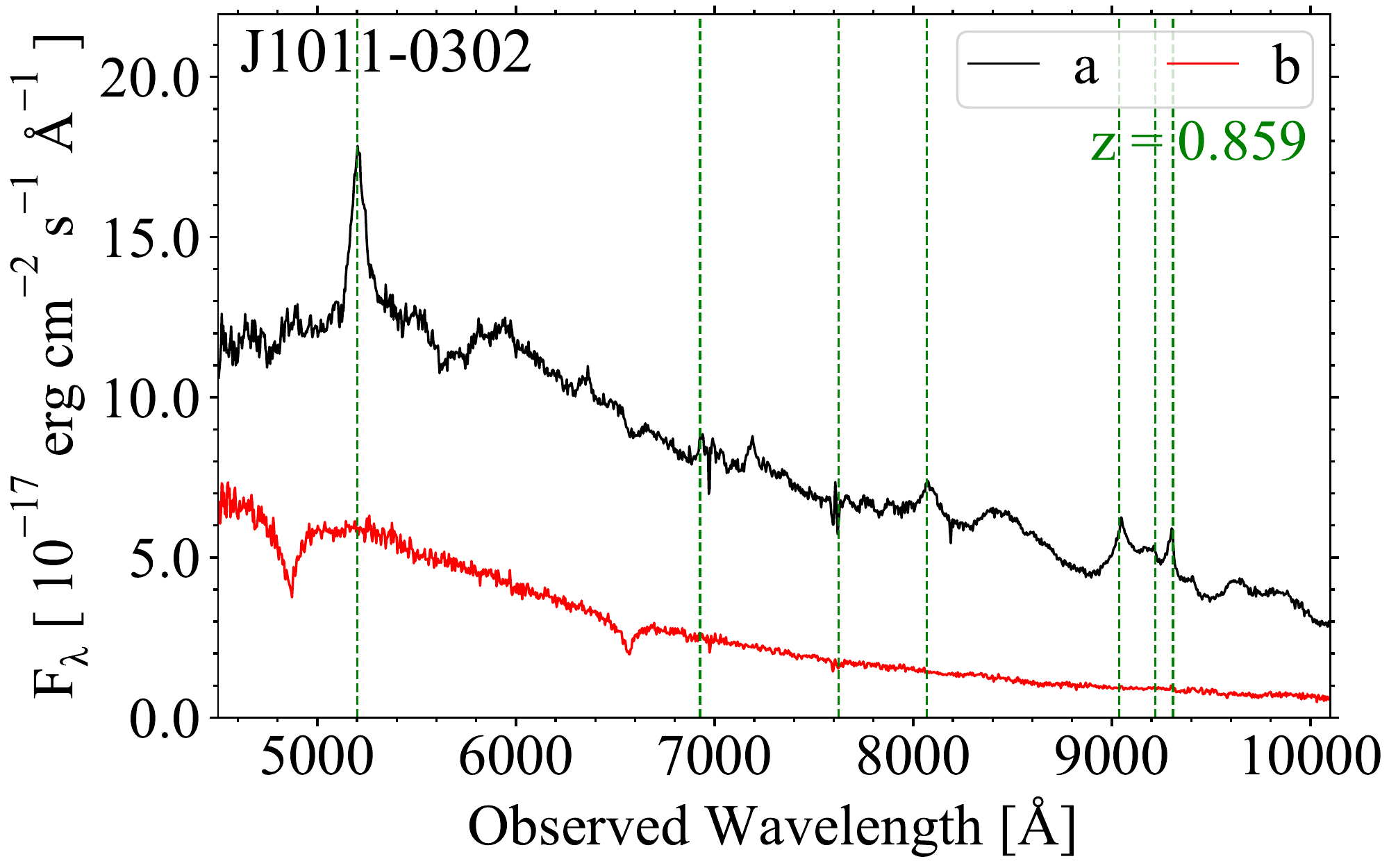}
	\includegraphics[width=0.33\textwidth]{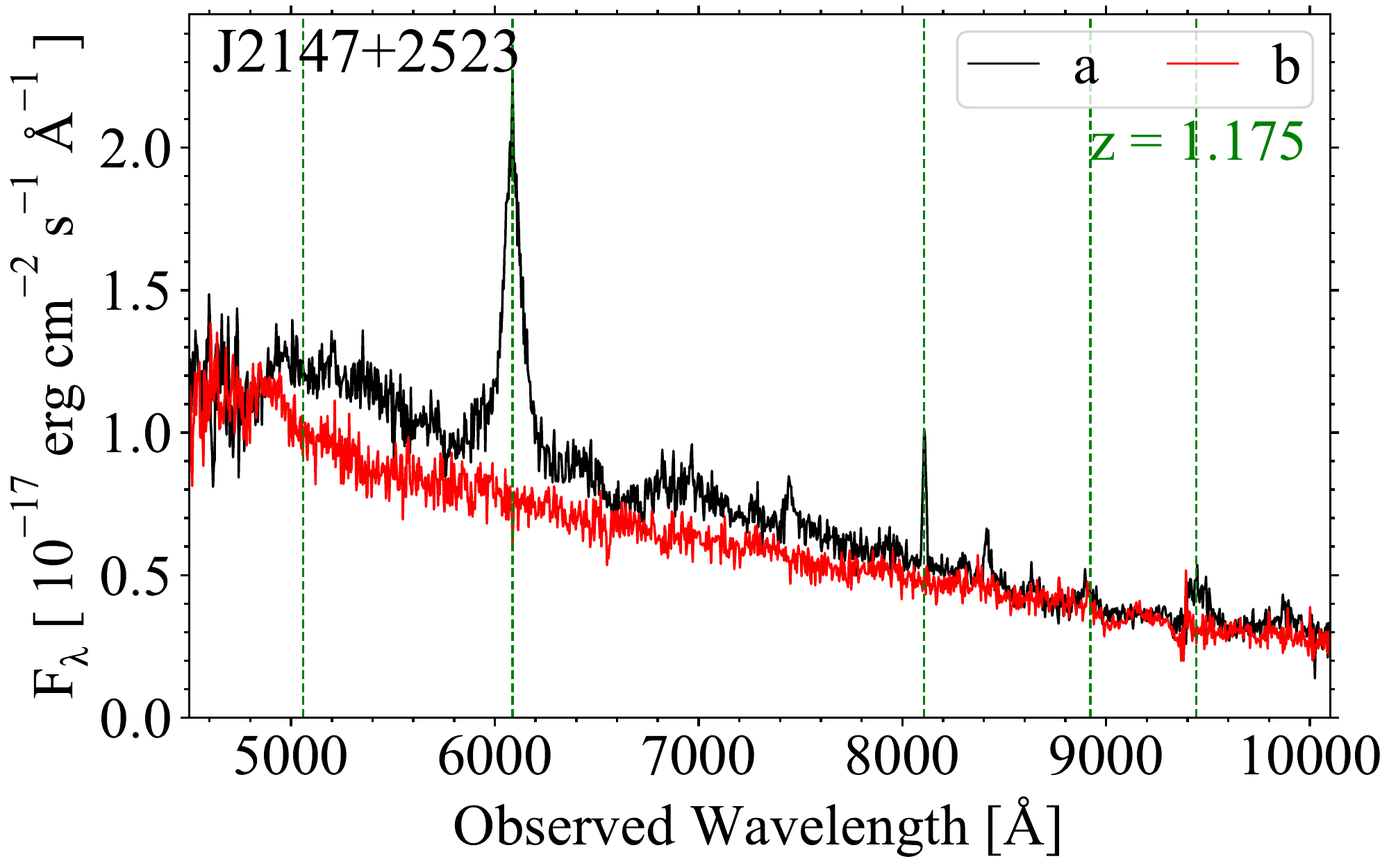}
	\includegraphics[width=0.33\textwidth]{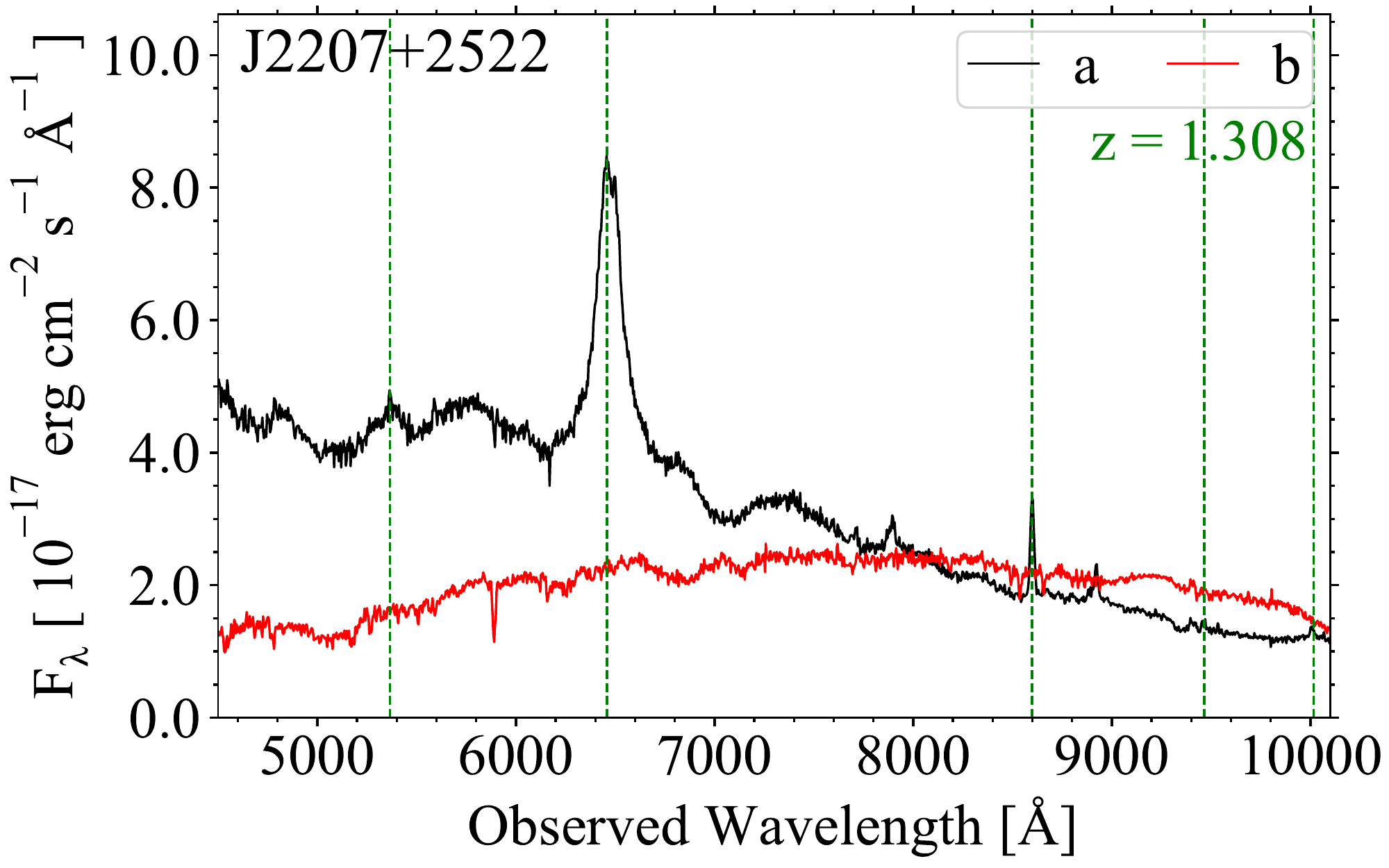}
	\includegraphics[width=0.33\textwidth]{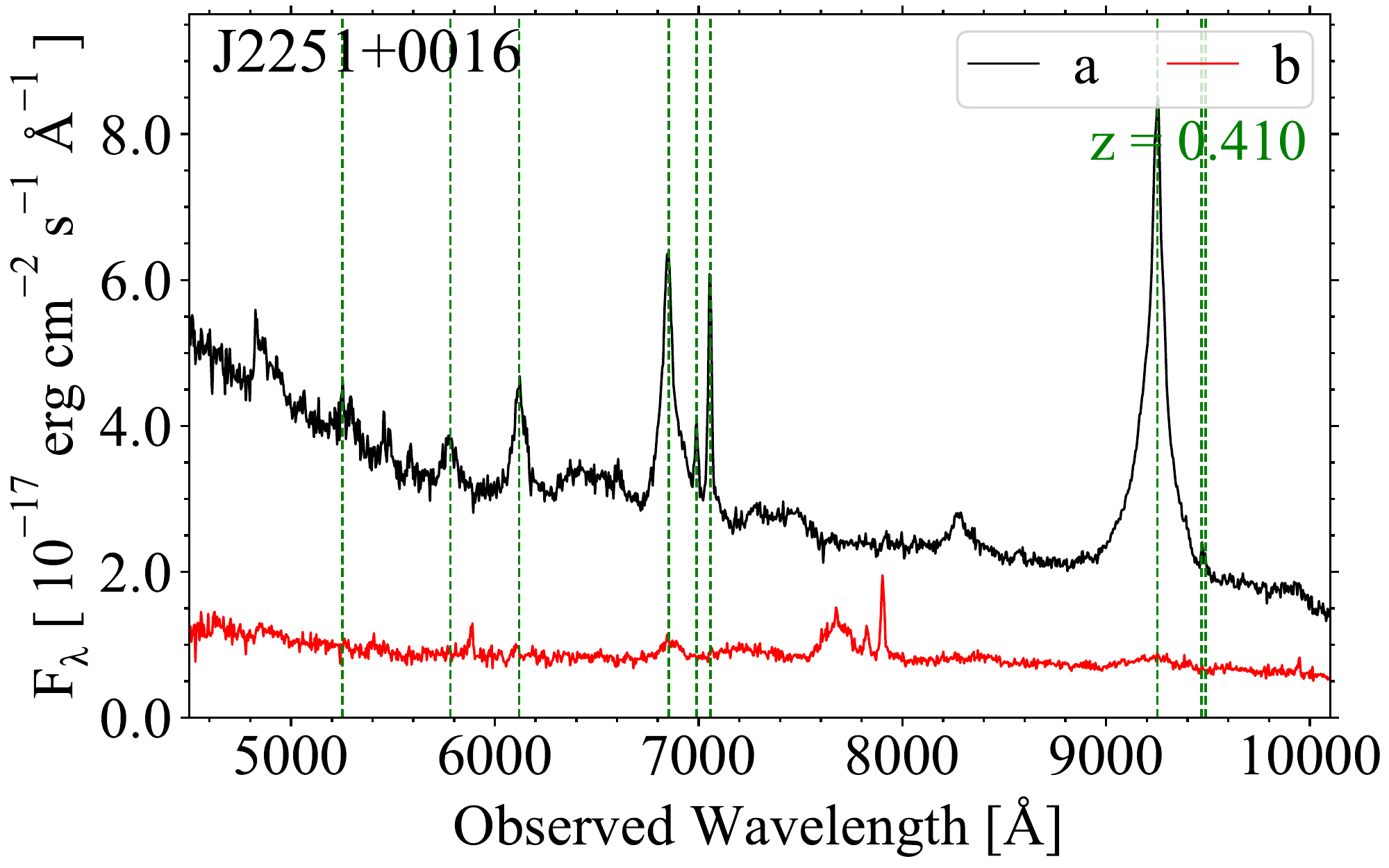}
	
    \caption{Spatially resolved Gemini/GMOS spectra of the 8 double quasar candidates. The black lines represent source a and the red lines represent source b. The green vertical dashed lines mark the expected quasar emission lines for source a.}
    \label{fig:gemini_spectra}
\end{figure*}

\subsection{DECaLS images and Gemini GMOS spectra}

I present the spatially resolved optical images of the eight targets from DECaLS and the decomposed optical spectra from Gemini. \autoref{fig:decals_images} shows the DECaLS $grz$ color composite images of the 8 double quasar candidates. Each source is detected in the DECaLS pipeline. The optical morphology are consistent with the two Point Spread Functions (PSFs) except for J2251+0016 at $z=0.410$, which has the extended emission from host galaxy. The coordinates and magnitudes of each source are listed in \autoref{tab:targets}. 


The spatially resolved Gemini/GMOS optcial spectra of the 8 double quasar candidates are shown in \autoref{fig:gemini_spectra} with the detail of the observations listed in \autoref{tab:targets}. The spatial profiles of the 8 targets are consistent with two Gaussian components in the spectra. The black and red lines represent the spectra of source a and b, respectively. The typical quasar emission lines for the primary source are marked in green vertical dashed lines. With the spatially-resolved spectra, I classify individual systems in details in \autoref{sec:classification} and derive the physical properties of the four double quasars in \autoref{sec:physical_properties}. 

\subsection{Classification of individual system}

\label{sec:classification}
Out of the eight observed targets, there are four double quasars (J0037+2058, J0118$-$0104, J0818+0601, and J0932+0722), one quasar pair at different redshifts (J2251+0016),  and three star-quasar superpositions (J1011$-$0302, J2147+2523, and J2207+2522). I provide my best-effort classifications for the individual systems below. I use velocity offsets between two sources, shapes of the emission lines, and other evidences from the literature to tell dual quasars and lensed quasars apart. Dual quasars are used throughout the paper for physical associated quasar pairs at same redshifts.

\subsubsection{Double quasar (dual/lensed quasar)}

\begin{itemize}
    \item \textbf{J0037+2058}: The broad \CIV\ and \MgII\ emission lines in both spectra confirm that both sources are quasars at $z=2.047$. Though two spectra are similar, the slopes of the continuum and the intensity of the emission lines such as \CIV\ are slightly different. It is likely that J0037+2058 is a dual quasar. However, the lensing hypothesis can not be excluded; the different flux ratio might be due to microlesing or dust extinction in the lens galaxy \citep[e.g.][]{Falco1999}.
    \item \textbf{J0118$-$0104}: The broad \MgII\ emission lines in both spectra confirm both sources are quasars at $z=0.739$. The shapes of emission lines between two sources are totally different and there is a noticeable velocity offset, both of which affirm that it is a dual quasar. A recent lens survey \citep{Lemon20} also reported it as a nearly identical quasar pair, not a lensed quasar.
    \item \textbf{J0818+0601}: The broad \CIV\ and \MgII\ emission lines in both spectra confirm that both sources are quasars at $z=2.363$. \citet{More16} classified it as a binary quasar because of the non-detection of the lens galaxy, however, recent spectropolarimetric observations identified it as a gravitationally lensed broad absorption line quasar \citep{Hutsemekers2020}. The absorption features are likely related to a lensed galaxy at $z=1.0$.
    \item \textbf{J0932+0722}: The broad \CIV\ and \MgII\ lines in both spectra confirm that both source are quasars at $z=1.996$ and $z=1.989$. The significant velocity offset of 850 km/s between two sources measured from \MgII\ lines affirm that it is a dual quasar, not a lens system. The flux ratio in optical is also significantly different than that in Infrared \citep{Inada08}. 
\end{itemize}

\subsubsection{Double quasar at different redshifts}
\begin{itemize}
    \item \textbf{J2251+0016}: Although both sources show broad \halpha\ and \hbeta\ emission lines, they have different redshifts at $z=0.410$ and $z=0.577$. It confirms that they are superposition of two quasars at different redshifts. 
\end{itemize}

\subsubsection{Star-Quasar superposition}

\begin{itemize}
    \item \textbf{J1011$-$0302:} The strong \halpha, \hbeta\ and H$\gamma$ absorption lines and the steep continuum in the spectrum of source b affirm that it is likely a F type star. 
    \item \textbf{J2147+2523:} Although two spectra have similar continuum slopes, the spectrum of source b does not show any emission lines. The weak \halpha\ and \hbeta\ absorption features as well as the G band (CH) absorption lines at $\sim$ 4300\AA\ in the spectrum of source b indicate that it is a G type star.
    \item \textbf{J2207+2522:} The \NaI\ $\lambda\lambda$ 5890,5896 \AA\ absorption lines as well as the \CaII\ $\lambda\lambda$ 8498,8542,8662 \AA\ lines are seen in the spectrum of source b. Based on the continuum shape and the absorption features, it is likely that source b is a K type star.
    
\end{itemize}

\subsection{Physical properties of double quasars}
\label{sec:physical_properties}

\begin{figure*}
	\includegraphics[width=0.48\textwidth]{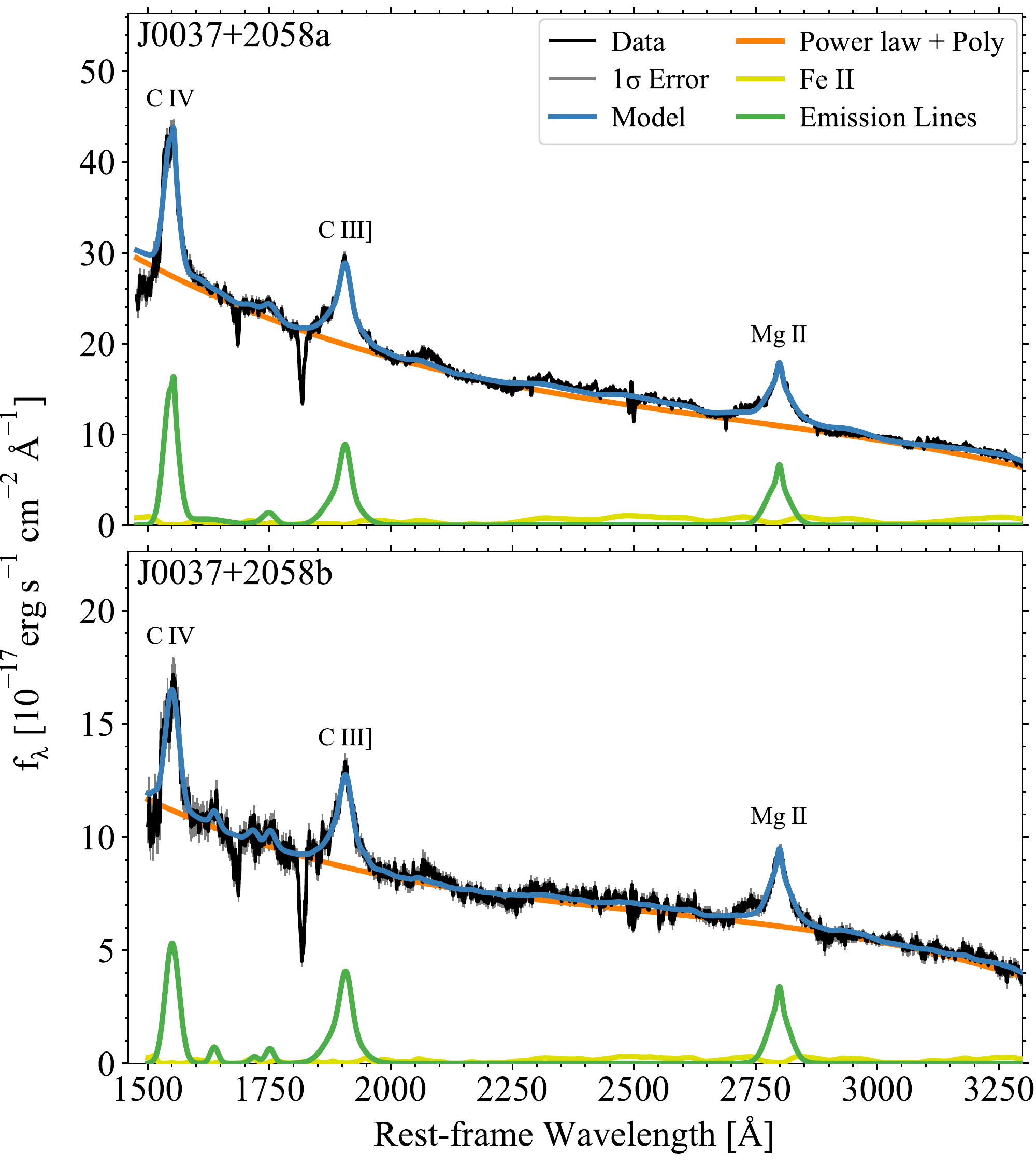}
	\includegraphics[width=0.48\textwidth]{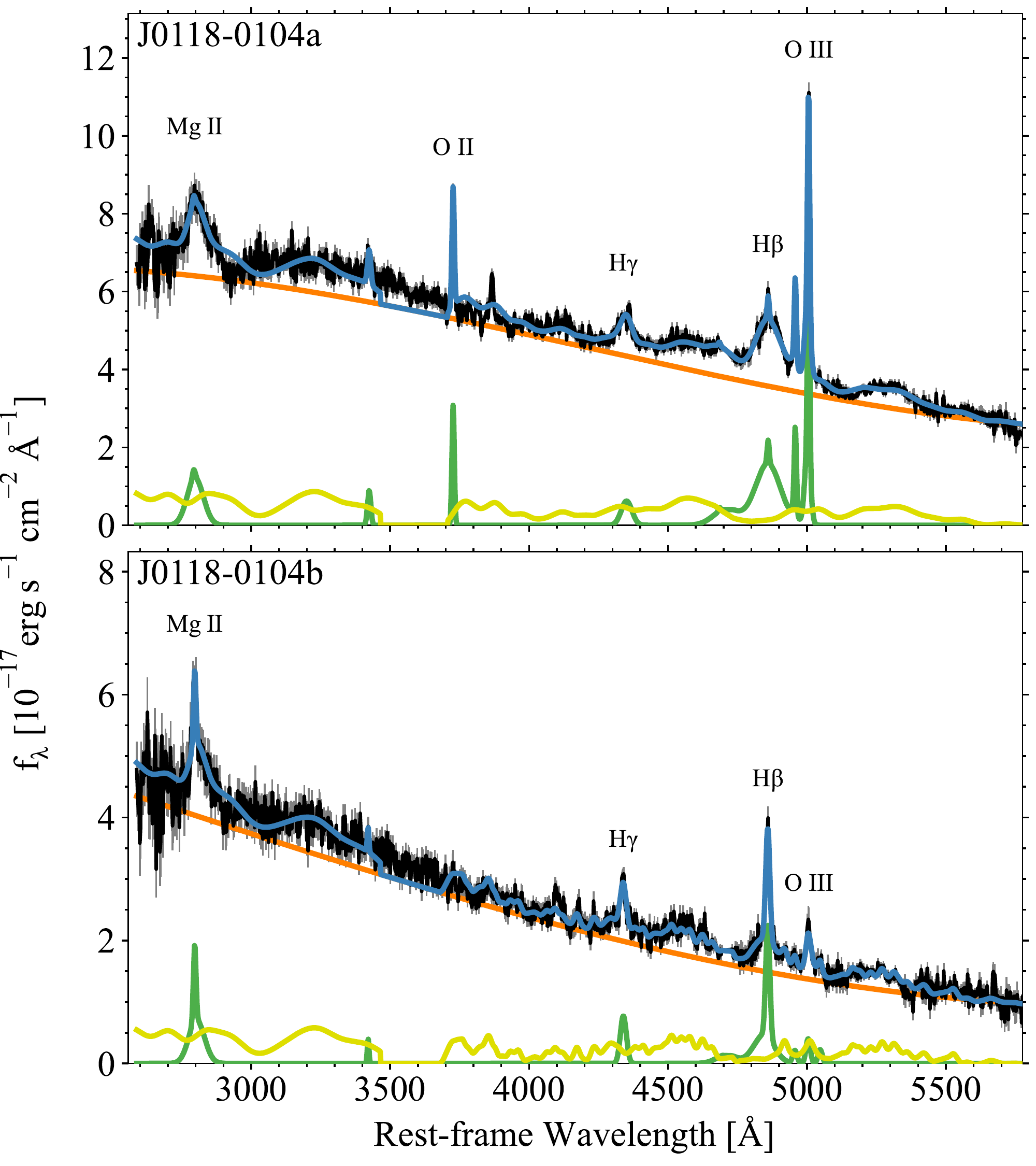}
	\includegraphics[width=0.48\textwidth]{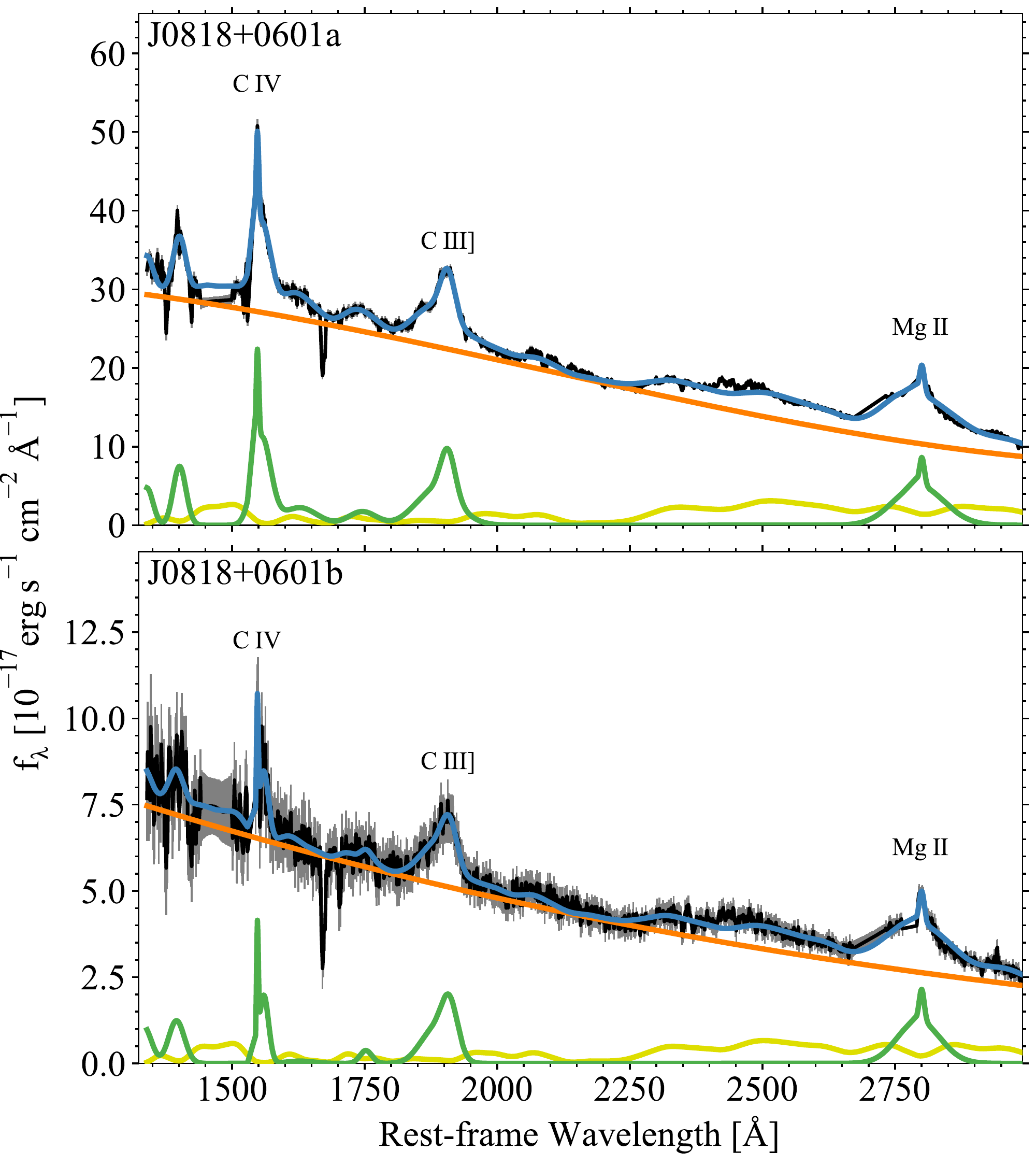}
	\includegraphics[width=0.48\textwidth]{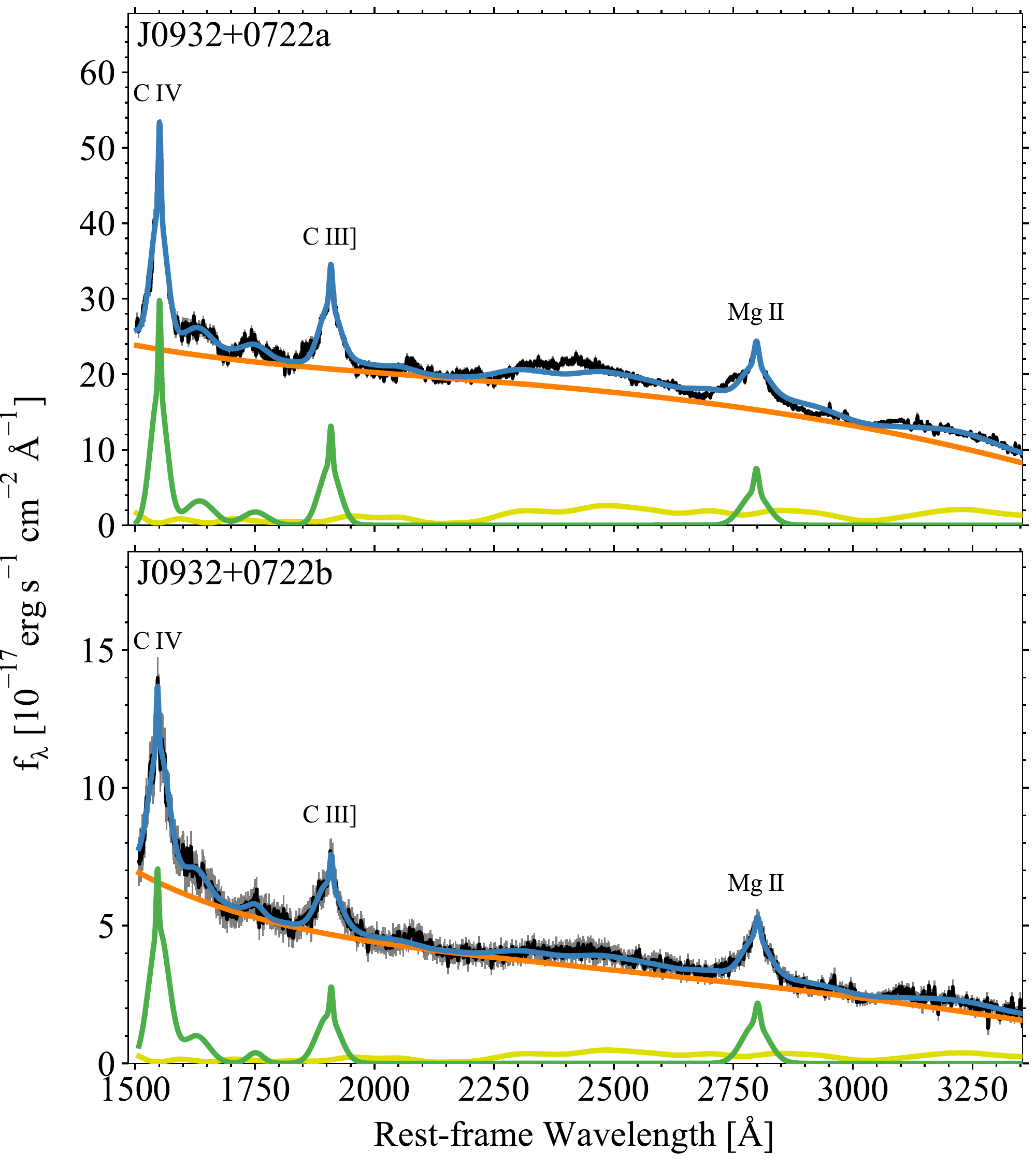}
    \caption{Spectral fitting for the four double quasars. Shown are the data (black), the 1$\sigma$ rms error (gray), the best-fit model (blue), the Fe II pseudo-continuum (yellow), the power-law plus polynomial model for the emission-line- and \FeII-subtracted continuum (orange), and the emission lines (green).}
    \label{fig:spectral_fitting}
\end{figure*}

\begin{table*}
 \caption{ Physical quantities of the four double quasars. 
  Listed from left to right are source names in J2000 coordinates, angular separations, projected physical separations, velocity offsets between two sources, FWHM of the broad \MgII\ lines, intrinsic luminosities at rest-frame 3000 \AA, BH masses, and Eddington ratios. }
 \label{tab:phys_properties}
 \begin{tabular}{cccccccc}
  \hline\hline
  Object & Sep. & $r_{\rm p}$ & $\Delta$v & FWHM$_{{\rm Mg}{\,\textsc{ii}}}$ & $\lambda L_{\lambda,3000}$ & log(M$_{\rm BH}$/ M$_{\odot}$)& $L_{\rm bol}$/$L_{\rm Edd}$ \\
  & (arcsec) & (kpc) & (km s$^{-1}$) & (km s$^{-1}$) & (ergs s$^{-1}$) & &   \\
  \hline
  J0037+2058a & 1.02 & 8.5 & 13$\pm$111 & 4013$\pm$160 & $8.58\times10^{45}$ & 9.03$\pm$0.03 & 0.35$\pm$0.08  \\
  J0037+2058b &  & &  & 3932$\pm$254 & $4.88\times10^{45}$ & 8.90$\pm$0.06  & 0.28$\pm$0.07 \\
  \hline
  J0118$-$0104a & 1.74 & 12.7 & 78$\pm$53 & 6447$\pm$1120 & $4.63\times10^{44}$ & 8.81$\pm$0.15 &  0.03$\pm$0.01\\
  J0118$-$0104b &  &  & & 1710$\pm$806 & $2.78\times10^{44}$ & 7.55$\pm$0.41 &  0.35$\pm$0.34 \\
  \hline
  J0818+0601a & 1.15 & 9.4 & 32$\pm$106 & 5661$\pm$276 & $1.12\times10^{46}$ & 9.39$\pm$0.04 & 0.20$\pm$0.05 \\
  J0818+0601b &  &  & & 4656$\pm$1738 & $2.90\times10^{45}$ & 8.93$\pm$0.32 & 0.15$\pm$0.12 \\
  \hline
  J0932+0722a & 1.33  & 11.1 & 641$\pm$113 & 3310$\pm$243 & $1.13\times10^{46}$ & 8.93$\pm$0.06 &  0.60$\pm$0.15 \\
  J0932+0722b &  &  & & 4393$\pm$676 & $2.05\times10^{45}$  &8.80$\pm$0.13 &  0.14$\pm$0.05\\
  \hline
 \end{tabular}
\end{table*}

To obtain the physical properties of each quasar for the four double quasars, I fit the individual spectra with PyQSOFit. PyQSOFit is a python fitting code to measure spectral properties of quasars \citep{PyQSOFit,Shen2019}. The spectral model consists of a pseudo-continuum, which is made from a power-law plus a third-order polynomial and the Fe {\sc II} emission templates \citep{Boroson1992,Vestergaard2001}, and a (or multiple) Gaussian component(s) for the narrow and broad emission lines. 
More specifically, I model \MgII\ and \CIV\ with one Gaussian for the narrow line component (defined as having a FWHM $<$1200 km s$^{-1}$) and up to two Gaussians for the broad-line component (defined as having a FWHM $\geq$1200 km s$^{-1}$). I remove the absorption features that are close to the emission lines and replace them with the linear interpolations using nearby pixels. The spectral fitting results are resented in \autoref{fig:spectral_fitting} and the few key measurement such as FWHM and continuum luminosities are listed in \autoref{tab:phys_properties}.

I estimate the BH mass of each quasar using the single-epoch virial BH mass estimators \citep{Shen2013}. The BH mass is estimated by
\begin{equation}
    \text{log}_{10} \bigg(\frac{M_{\text{BH}}}{M_{\odot}}\bigg) = a + b \text{ log}_{10}\bigg(\frac{\lambda L_{\lambda}}{10^{44}\,\text{erg s}^{-1}}\bigg) + 2 \text{ log}_{10}\bigg(\frac{\text{FWHM}}{\text{km s}^{-1}}\bigg),
\end{equation}
where $L_\lambda$ is the monochromatic continuum luminosity at wavelength $\lambda=$3000\AA\ , FWHM is the full width at half maximum of the \MgII\ line, and $a=0.86$ and $b=0.5$ are the coefficients derived from \citet{Vestergaard2009}. I choose the \MgII\ lines as the BH mass estimator given that the monochromatic continuum luminosity at wavelength $\lambda=$1350\AA\ are not available in most spectra and the \CIV\ lines are too close to the edge. \CIV\ is also often subject to outflows and have larger scatter in comparison to \hbeta\ masses \citep{Shen2012}.  

I estimate the bolometric luminosity from the monochromatic luminosity at 3000\AA\ using bolometric correction values from \citet{richards06}. The bolometric luminosity under the assumption of isotropy is  
\begin{equation}
    L_{\rm bol} = \lambda L_{\lambda} \times BC,
\end{equation}
where $L_\lambda$ is the monochromatic continuum luminosity at rest-frame wavelength $\lambda$=3000 \AA\ and BC = 5.62 is the bolometric correction from 3000 \AA\ \citep{richards06}. The Eddington luminosity is derived by 
\begin{equation}
    L_{\rm Edd} \sim 1.26 \times 10^{38} (\frac{M_{\rm BH}}{M_{\odot}})\ { \rm erg s}^{-1}
\end{equation}
\autoref{tab:phys_properties} lists the physical properties of the four double quasars including BH masses, projected physical separations, intrinsic luminosity and Eddington ratios.

\section{Discussions}
\label{sec:discussions}

\subsection{Comparison to previous work and implications on future surveys}

\begin{figure}
	\includegraphics[width=\columnwidth]{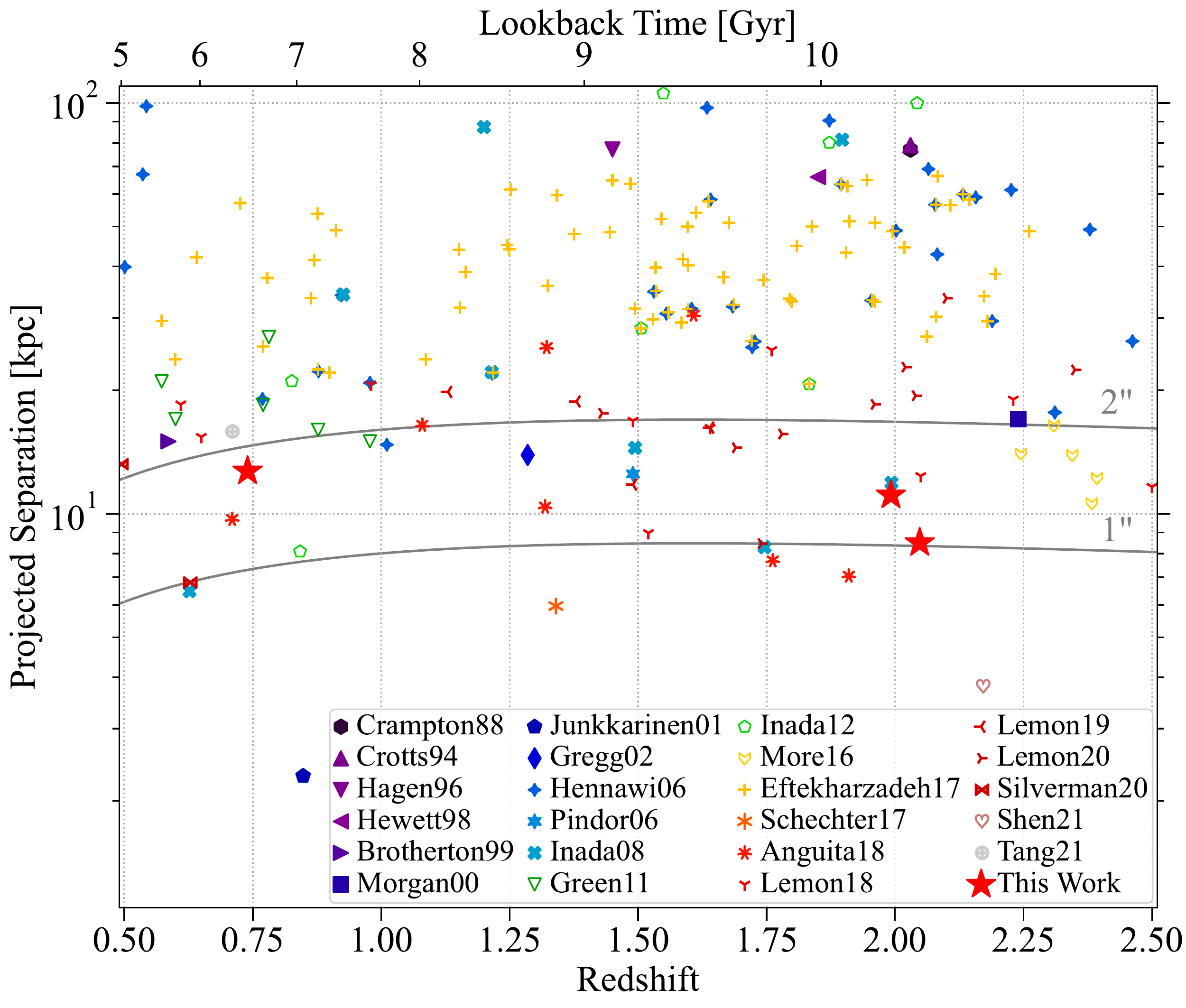}
	
    \caption{Redshifts and projected physical separations of the three confirmed dual quasars in comparison to the dual quasars at 0.5 $< z <$ 2.5 from the literature \citep{Crampton88,Crotts94,Hagen96,Hewett98,Brotherton99,Morgan00,Junkkarinen01,Gregg02,Hennawi06,Pindor06,Inada08,Green11,Inada12,More16,Eftekharzadeh17,Schechter17,Anguita18,Lemon18,Lemon19,Lemon20,Silverman20,Shen2021,Tang21}. A dual quasar is defined as two quasars with velocity difference $<$ 2000 km s$^{-1}$. The grey curves are the projected physical separations as a function of redshift with fixed angular separations of 1 arcsec and 2 arcsec. }
    \label{fig:redshift_rp}
\end{figure}

In the past few decades, significant progresses have been made to find dual quasar at high redshift (\autoref{fig:redshift_rp}). However, due to the resolution limits of ground-based optical surveys, most double quasars are at separation of $>2$ arcsec ($\gtrsim16$ kpc at $z>1$). The newly discovered double quasar with separation of $\sim10$ kpc increase the sample statistic in this incomplete region. It also proves that by combining the existing quasar catalogs with new imaging surveys, which have better data quality and classification pipelines, it is possible to search small-separation double quasars that were not found before.

It is practicable that the detection algorithm could identify blended pairs even if the separation is below the typical angular resolution. \citet{ChenYC2021b} recently reported tens of the sub-arcsec dual quasar candidates discovered in the Hubble Space Telescope. A significant fraction of those sub-arcsec pairs are identified as two sources in the DECaLS catalog, showing possible source identification even below angular resolution. Although I expect that the completeness of source identification should drop drmatically in the sub-arcsec regime, identifying those pair at smaller scale will benefit our understanding in the quasar clustering and possible merger-induced enhancement in quasar activity.

New imaging surveys such as the Dark Energy Survey (DES), which has excellent image quality ($r$-band PSF FHWM of 0\farcs96; \citealt{DESDR1}) and covers the southern hemisphere, will explore the untouched region of sky; in synergy with the ongoing all-sky spectroscopic surveys like SDSS-V \citep{SDSSV}, they will increase the number of small-scale double quasars significantly. Future all-sky imaging surveys like Legacy Survey of Space and Time (LSST) with Vera C. Rubin observatory will push the limit much further for dual quasars with less masses and smaller separations given its remarkable 10-yr depth of $\sim27$ and typical seeing of 0\farcs63 \citep{LSST}. The future Nancy Grace Roman Space Telescope \citep{Spergel2015,Akeson2019}, which has angular resolution like Hubble Space Telescope but is hundred times more powerful in survey speed, will provide a remarkable sample that is unable to achieve on the ground. By expanding the parameter space in mass, separation, and redshift, we will be able to explore evolution and assembly of SMBHs in space and time in details.

\subsection{Number of dual quasar as a function of projected separation}

\begin{figure}
	\includegraphics[width=\columnwidth]{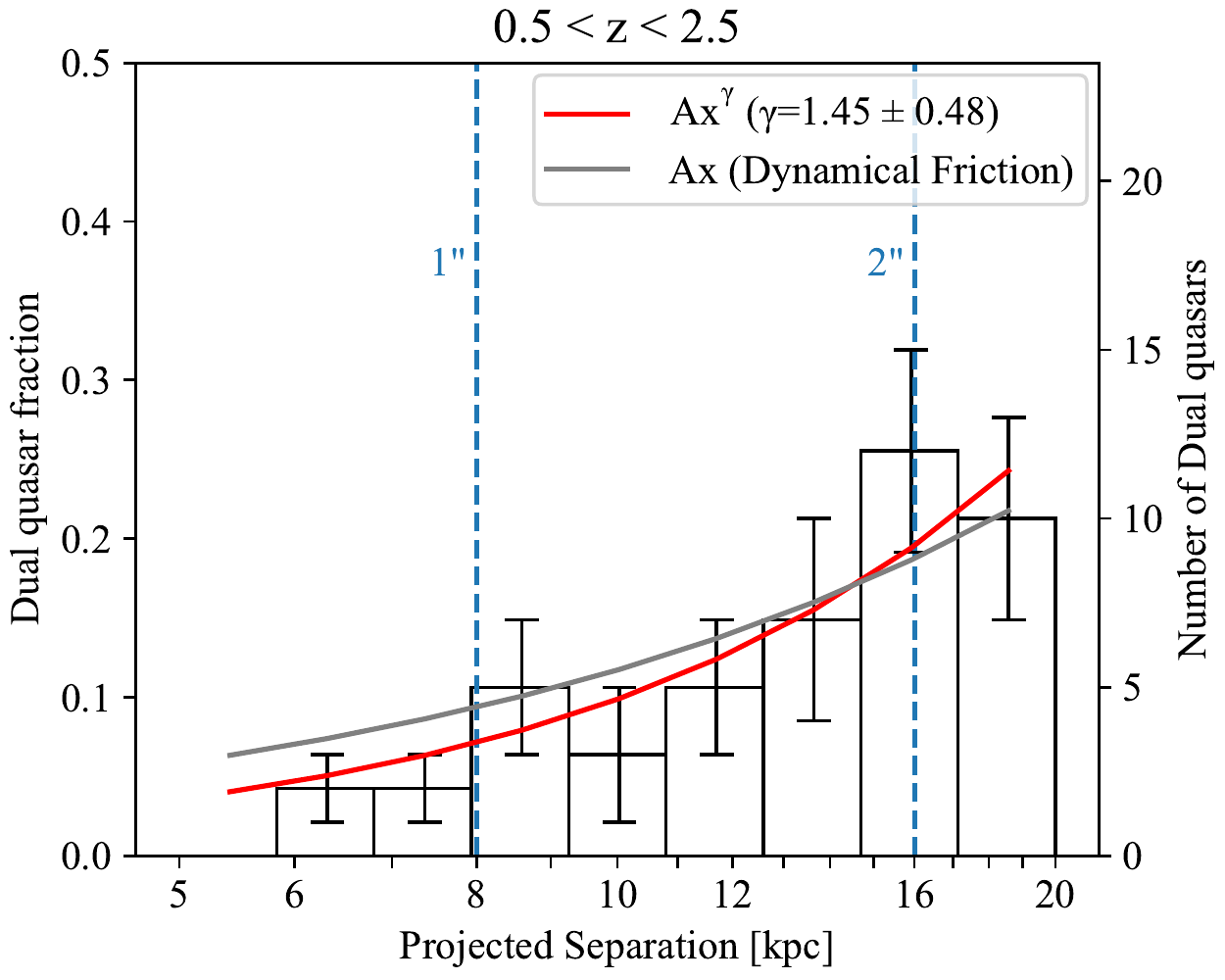}
	
    \caption{Number of dual quasars or dual quasar fraction as a function of projected separation at 0.5 $< z <$ 2.5. The blue vertical dashed lines mark the projected physical separation for angular separation of 1 arcsec and 2 arcsec at $z=1$. The red curve is the best-fit curve of the power-law Ax$^{\gamma}$ for the data above 1 arcsec and the grey curve is the best-fit curve of the linear relation ($\gamma$=1) expected from dynamical friction.}
    \label{fig:dual_frac}
\end{figure}


The demographics of dual quasar yield fundamental constraints on models of quasar evolution and the assembly of SMBHs \citep{Hennawi06,Hopkins2007,daAngela2008}. \autoref{fig:dual_frac} shows the number of dual quasars or the dual quasar fraction as a function of projected separation for dual quasar with separation between 5 kpc and 20 kpc at $0.5 < z < 2.5$ compiled from this work and the literature \citep{Brotherton99,Morgan00,Gregg02,Hennawi06,Pindor06,Inada08,Green11,Inada12,More16,Schechter17,Anguita18,Lemon18,Lemon19,Lemon20,Silverman20,Tang21}. I assume that J0037+2058 is a dual quasar and select all the double quasars that are not lensed quasars from the literature. The sample below 1 arcsec ($\sim$7.5 kpc at $z=1$) is highly incomplete because of the stringent angular resolution of ground-based surveys. I fit a power-law relation $\rm Ax^{\gamma}$ to the data points above 1 arcsec by removing the first three bins. Below tens kpc, the dynamical evolution of dual BHs is determined by the dynamical friction; the dynamical friction timescale is proportional to the BH separation \citep{begelman80,Yu2002}. If the quasar duty cycle does not change during merger, the number of dual quasars should be inversely proportional to separation. The power-law coefficient $\gamma$ of 1.45 from my fit is slightly larger than unity, which is expected from dynamical friction. The slopes are consistent within 1$\sigma$ uncertainty when I use different bin sizes. The steeper index is likely caused by the incompleteness of dual quasars at $<$ 15 kpc, although gas drag in gas-rich mergers at high redshift or enhanced quasar activity due to merger could alter the slope significantly. Finding more small-scale dual quasars will provide a more accurate estimation of the distribution. It should also be noted that the collection of dual quasars is from various surveys and could be heterogeneous. Besides, few dual quasars might turn out to be lensed quasars due to the lack of deep IR observations to detect lens galaxies. A detailed analysis with more complete or homogeneous sample at small scale is needed to understand the systematics better and determine the underlying physical principles.

\subsection{Color selection efficiency}

I demonstrate the high efficiency of selecting double quasars using the color selection made from $grz$-bands photometry. Among the 34 double quasar candidates passing the color criteria, 26 are observed in the Gemini program or reported in the literature \citep{Inada08,Inada10,Kayo10,Inada12,Jackson12,More16}. There are only 5 ($\sim$19\%) that are star-quasar superpositions, which suggests that the chance positions with stars can be significantly reduced by using the color-color diagram. My selection is only based on few simple straight line cuts that might exclude double quasars close to the stellar locus. I also find that the colors of the three confirms stars in my Gemini observations are far from those of most quasars (\autoref{fig:color}).  A more sophisticated selection criteria (e.g., distance from the stellar locus or color differences of two sources in a pair) will recover more double quasars.

\section{Conclusions}
\label{sec:conclusions}

I observe eight small-scale double quasar candidates with Gemini long-slit optical spectroscopy. Those double quasar candidates are selected from DECaLS with separation of 1$-$2 arcsec (\autoref{fig:decals_images}). The spatially resolved spectra reveal four double quasars, in which one is newly discovered. Two of the four double quasars (J0118-0104 and J0932+0722) are identified as physical quasar pairs based on the shapes of emission lines and velocity offsets, while J0037+2058 could be either a dual quasar or a lensed quasar because of the nearly identical spectra between two sources and the small velocity offset. I also conduct the spectral fitting for the four double quasars (\autoref{fig:spectral_fitting}) and report the physical properties of each source such as projected separation, BH masses, Eddington ratios (\autoref{tab:phys_properties}). 

I obtain a steeper power-law index of 1.45 for the number of dual quasar as a function of projected separation using the dual quasar sample compiled from this work and the literature. The power-law coefficient is higher than that expected from the dynamical friction, which is likely due to the incomplete sample of dual quasars at $<15$ kpc. Detailed study on the complete and homogeneous sample is needed to determine the accurate distribution.

A low contamination rate ($\sim$19\%) of star-quasar superpositions demonstrates the high efficiency of selecting double quasars with color criteria, though a more sophisticated selection will reveal more hidden double quasars. The newly discovered double quasar with $\sim$10 kpc not only increases the sample statistics in the incomplete part of the redshift-separation regime (\autoref{fig:redshift_rp}) but also shows the possibility of finding more double quasars by combining new imaging surveys with existing quasar catalogs. New and future imaging surveys like DES and LSST will lead to more discoveries for double quasars at kpc scale at high redshift.

\section*{Acknowledgements}


YCC thanks Xin Liu, Hengxiao Guo, Colin Burke, and Qian Yang for discussions and comments. YCC thank Trent Dupuy, Julia Scharwaechter, Kristin Chiboucas, and Jason Chu for their helps with scheduling the Gemini observations. YCC acknowledges support from the University of Illinois Campus Research Board and NSF grant AST-2108162. YCC is supported by the government scholarship to study aboard from the ministry of education of Taiwan and by the graduate fellowship from the Center for AstroPhysical Surveys at the University of Illinois’ National Center for Supercomputing Applications.

Based on observations obtained at the international Gemini Observatory, a program of NSF's NOIRLab, which is managed by the Association of Universities for Research in Astronomy (AURA) under a cooperative agreement with the National Science Foundation on behalf of the Gemini Observatory partnership: the National Science Foundation (United States), National Research Council (Canada), Agencia Nacional de Investigaci\'{o}n y Desarrollo (Chile), Ministerio de Ciencia, Tecnolog\'{i}a e Innovaci\'{o}n (Argentina), Minist\'{e}rio da Ci\^{e}ncia, Tecnologia, Inova\c{c}\~{o}es e Comunica\c{c}\~{o}es (Brazil), and Korea Astronomy and Space Science Institute (Republic of Korea).

This work was enabled by observations made from the Gemini North telescope, located within the Maunakea Science Reserve and adjacent to the summit of Maunakea. I are grateful for the privilege of observing the Universe from a place that is unique in both its astronomical quality and its cultural significance.

This research uses services or data provided by the Astro Data Lab at NSF's NOIRLab. NOIRLab is operated by the Association of Universities for Research in Astronomy (AURA), Inc. under a cooperative agreement with the National Science Foundation.

The Legacy Surveys consist of three individual and complementary projects: the Dark Energy Camera Legacy Survey (DECaLS; Proposal ID \#2014B-0404; PIs: David Schlegel and Arjun Dey), the Beijing-Arizona Sky Survey (BASS; NOAO Prop. ID \#2015A-0801; PIs: Zhou Xu and Xiaohui Fan), and the Mayall z-band Legacy Survey (MzLS; Prop. ID \#2016A-0453; PI: Arjun Dey). DECaLS, BASS and MzLS together include data obtained, respectively, at the Blanco telescope, Cerro Tololo Inter-American Observatory, NSF's NOIRLab; the Bok telescope, Steward Observatory, University of Arizona; and the Mayall telescope, Kitt Peak National Observatory, NOIRLab. The Legacy Surveys project is honored to be permitted to conduct astronomical research on Iolkam Du'ag (Kitt Peak), a mountain with particular significance to the Tohono O'odham Nation.

NOIRLab is operated by the Association of Universities for Research in Astronomy (AURA) under a cooperative agreement with the National Science Foundation.

This project used data obtained with the Dark Energy Camera (DECam), which was constructed by the Dark Energy Survey (DES) collaboration. Funding for the DES Projects has been provided by the U.S. Department of Energy, the U.S. National Science Foundation, the Ministry of Science and Education of Spain,  the Science and Technology Facilities Council of the United Kingdom, the Higher Education Funding Council for England, the National Center for Supercomputing Applications at the University of Illinois at Urbana-Champaign, the Kavli Institute of Cosmological Physics at the University of Chicago, the Center for Cosmology and Astro-Particle Physics at the Ohio State University, the Mitchell Institute for Fundamental Physics and Astronomy at Texas A\&M University, Financiadora de Estudos e Projetos, Funda{\c c}{\~a}o Carlos Chagas Filho de Amparo {\`a} Pesquisa do Estado do Rio de Janeiro, Conselho Nacional de Desenvolvimento Cient{\'i}fico e Tecnol{\'o}gico and the Minist{\'e}rio da Ci{\^e}ncia, Tecnologia e Inova{\c c}{\~a}o, the Deutsche Forschungsgemeinschaft and the Collaborating Institutions in the Dark Energy Survey. The Collaborating Institutions are Argonne National Laboratory, the University of California at Santa Cruz, the University of Cambridge, Centro de Investigaciones Energ{\'e}ticas, Medioambientales y Tecnol{\'o}gicas-Madrid, the University of Chicago, University College London, the DES-Brazil Consortium, the University of Edinburgh, the Eidgen{\"o}ssische Technische Hochschule (ETH) Z{\"u}rich, Fermi National Accelerator Laboratory, the University of Illinois at Urbana-Champaign, the Institut de Ci{\`e}ncies de l'Espai (IEEC/CSIC), the Institut de F{\'i}sica d'Altes Energies, Lawrence Berkeley National Laboratory, the Ludwig Maximilians Universitat Munchen and the associated Excellence Cluster Universe, the University of Michigan, NSF's NOIRLab, the University of Nottingham, the Ohio State University, the University of Pennsylvania, the University of Portsmouth, SLAC National Accelerator Laboratory, Stanford University, the University of Sussex, and Texas A\&M University.

The Legacy Survey team makes use of data products from the Near-Earth Object Wide-field Infrared Survey Explorer (NEOWISE), which is a project of the Jet Propulsion Laboratory/California Institute of Technology. NEOWISE is funded by the National Aeronautics and Space Administration.

The Legacy Surveys imaging of the DESI footprint is supported by the Director, Office of Science, Office of High Energy Physics of the U.S. Department of Energy under Contract No. DE-AC02-05CH1123, by the National Energy Research Scientific Computing Center, a DOE Office of Science User Facility under the same contract; and by the U.S. National Science Foundation, Division of Astronomical Sciences under Contract No. AST-0950945 to NOAO.

\section*{Data Availability}

The Gemini data will be available in the Gemini Observatory Archive at \url{https://archive.gemini.edu/searchform} after the proprietary period. The DECaLS data are available at \url{https://www.legacysurvey.org}.

\bibliographystyle{mnras}
\bibliography{ref_new} 

\end{document}